\documentclass[%
 reprint,
nobibnotes,
 amsmath,amssymb,
 aps,
]{revtex4-1}
\pdfoutput=1
\usepackage{bm}
\usepackage{xcolor}
\usepackage{graphicx}
\usepackage[utf8]{inputenc}
\usepackage{epstopdf}
\usepackage{afterpage}
\usepackage{setspace}
\normalsize
\usepackage{booktabs}
\usepackage{hyperref}

\usepackage{placeins}
\usepackage{tabularx}

\usepackage[framemethod=TikZ]{mdframed}

\usepackage{subcaption}

\expandafter\ifx\csname package@font\endcsname\relax\else
 \expandafter\expandafter
 \expandafter\usepackage
 \expandafter\expandafter
 \expandafter{\csname package@font\endcsname}%
\fi
\usepackage{xcolor}
\definecolor{lightblue}{RGB}{63, 72, 138}

\begin{document}
\mdfsetup{frametitlealignment=\centering} 
\mdfdefinestyle{mystyle}{roundcorner=4pt,innerleftmargin=0.5cm,innerrightmargin=0.5cm,innertopmargin = 0.3cm,innerbottommargin = 0.3 cm, frametitlebackgroundcolor=lightblue!33,linecolor=lightblue}

\setstretch{1.0}
\title{Tutorial: How to Train a Neural Network Potential}

\author{Alea Miako Tokita}
\email{alea.tokita@ruhr-uni-bochum.de}
\author{J\"{o}rg Behler}
\email{joerg.behler@ruhr-uni-bochum.de}
\affiliation{Lehrstuhl f\"ur Theoretische Chemie II, Ruhr-Universit\"at Bochum, 44780 Bochum, Germany}
\affiliation{Research Center Chemical Sciences and Sustainability, Research Alliance Ruhr, 44780 Bochum, Germany}

\begin{abstract}
The introduction of modern Machine Learning Potentials (MLP) has led to a paradigm change in the development of potential energy surfaces for atomistic simulations. By providing efficient access to energies and forces, they allow us to perform large-scale simulations of extended systems, which are not directly accessible by demanding first-principles methods. In these simulations, MLPs can reach the accuracy of electronic structure calculations provided that they have been properly trained and validated using a suitable set of reference data. Due to their highly flexible functional form the construction of MLPs has to be done with great care. In this Tutorial, we describe the necessary key steps for training reliable MLPs, from data generation via training to final validation. The procedure, which is illustrated for the example of a high-dimensional neural network potential, is general and applicable to many types of MLPs.
\end{abstract}

\maketitle

\section{Introduction}\label{sec:introduction}

\begin{figure}[t]
    \centering
    \includegraphics[width=\columnwidth]{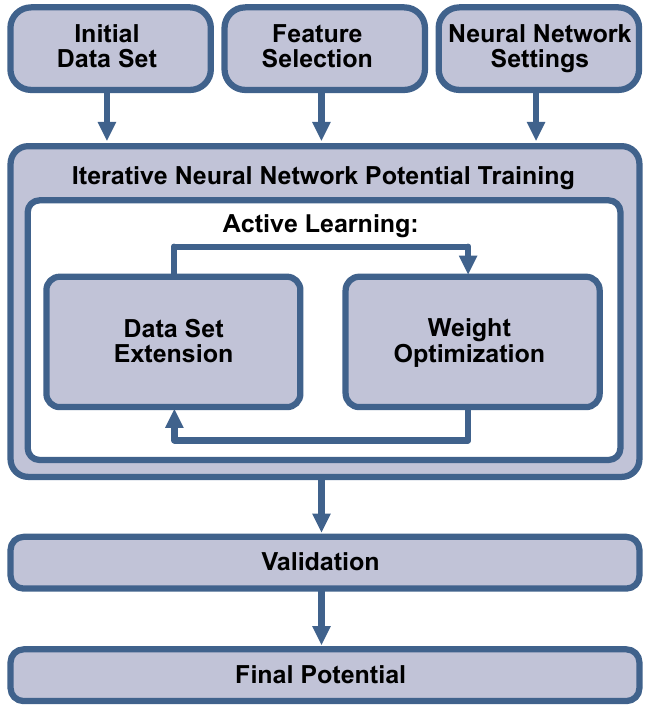}
    \caption{Overview of the construction of a neural network potential.}
    \label{fig:overview}
\end{figure}

In recent decades, advances in atomistic simulations have revolutionized the way of studying complex systems in many fields, from chemistry and physics via materials science to the life sciences. 
At the present time, computer simulations allow to understand complex experimental data, and to rationalize or even predict the properties of molecules and solids as well as their reactions based on detailed structural and dynamical information at the atomic level.
A fundamental requirement to perform such simulations is the knowledge about the atomic interactions, i.e., the potential energy and the forces, which in principle can be obtained by solving the Schr\"odinger equation. Unfortunately, such quantum mechanical calculations are computationally very demanding, even if relatively efficient methods such as density functional theory (DFT)~\cite{P3596,P3051} are used. Therefore, the accessible time and length scales of ab initio molecular dynamics (MD) simulations~\cite{P0433,B0007}, in which the energies and forces are determined by DFT for each visited atomic configuration, are limited to a few hundred atoms and tens to hundreds of picoseconds. 
\\
The computational effort could be substantially reduced if the multidimensional function defining the relation between the atomic positions and the potential energy, i.e., the potential energy surface (PES), would be directly accessible. Founded on the Born-Oppenheimer approximation of quantum mechanics~\cite{P3970}, the PES contains a wealth of information, such as global and local minima defining stable and metastable geometries, barriers of chemical reactions, and forces governing the nuclear motion. Approximate atomistic potentials and force fields representing the PES in analytic form have been used for decades~\cite{P0820,P0341,P1450,P0665,P1453,P1382,P1454}, and conventional approaches rely on physically reasonable simplifications to increase the computational efficiency. Such approximations typically reduce the accuracy but often still allow to maintain an acceptable transferability of the potential.
\\
In recent years, machine learning (ML) has emerged as a new and powerful computational tool with many applications in the chemical and physical sciences~\cite{P5779,B0109,P6530}, such as drug design~\cite{P0845,P6545}, synthesis planning~\cite{P5445,P5970}, protein structure prediction~\cite{P6544} and the analysis and prediction of spectra~\cite{P6107,P6130}. Another important usage of ML is the representation of the PES by machine learning potentials (MLP), which has first been proposed more than a quarter of a century ago~\cite{P0316}. Since then, MLPs have witnessed tremendous progress~\cite{P4885,P6121,P6102,P6112,P5793,P4263,P5673,P5788,P4444,P6131,P3033,P2559,P5977,P5877} by exploiting the flexibility of ML methods such as neural networks (NNs) to learn the atomic interactions from reference energies and forces obtained from electronic structure calculations. The resulting analytical ML expression can then provide the energy and its derivatives with about the accuracy of the reference method at a small fraction of the computational costs, which works well as long as the requested structures are not too different from those included in the training set. Due to this limited extrapolation capabilities, often large and structurally diverse data sets are used in the construction of MLPs. Another advantage of MLPs is their ability to represent all types of bonding, such as covalent and metallic bonds as well as ionic and dispersion interactions, at the same level of accuracy by utilizing general and unbiased functional forms. Moreover, like the underlying electronic structure methods, they are ``reactive'', i.e., they can describe the making and breaking of bonds. However, the computational costs of MLPs are usually higher than those of simple classical force fields, since a large number of terms needs to be evaluated and more complex coordinate transformations are required.  
\\
Current MLPs can be assigned to different generations~\cite{P6018,P5977}, and a suitable choice depends on the specific system of interest. While MLPs of the first generation are restricted to low-dimensional systems, the introduction of second-generation high-dimensional neural network potentials in 2007 paved the way to the application of MLPs to high-dimensional systems containing large numbers of atoms~\cite{P1174,P4106,P5128,P6018}. This has been achieved by expressing the potential energy of the system as a sum of atomic energies and by the introduction of atom-centered symmetry functions (ACSFs) as atomic environment descriptors maintaining the mandatory translational, rotational and permutational invariances of the energy~\cite{P2882}. Over the years, many types of second-generation MLPs have been introduced differing in the employed ML algorithms and descriptors, such as various forms of neural network potentials~\cite{P1174,P4945,P5366,P6017,P5596}, Gaussian approximation potentials \cite{P2630,P4429}, moment tensor potentials \cite{P4862}, spectral neighbor analysis potentials \cite{P4644}, atomic cluster expansion\cite{P5794} and many others. 
\\
Long-range electrostatic interactions are included in third-generation MLPs~\cite{P2962,P5577,P5313,P6200,P6347}. Here, the necessary charges or even multipoles can be predicted as a function of the atomic environments by machine learning~\cite{P2391,P2962,P5885,P5577}. Also MLPs with explicit dispersion interactions beyond the local atomic environments can be classified as third-generation~\cite{P5675,P6537,P6319}.
Fourth-generation MLPs~\cite{P4419,P5859,P5932,P5933,P6122}, which employ global charge equilibration techniques~\cite{P1448} or self-consistent charge distributions~\cite{P5859}, can describe non-local phenomena such as long-range charge transfer and can be applied to multiple charge states of a system. 
\\
While many MLPs make use of predefined features to describe environment-dependent atomic properties such as energies, charges, or electronegativities, in recent years different forms of message passing neural networks~\cite{P5368} have also been developed for the representation of PESs. Here, the description of the atomic environments is included in the training process by iteratively passing structural information through the system~\cite{P4937,P5817,P5577,P5366,P6026,P6017}, which in principle can provide information beyond a predefined local environment.
\\
Despite the significant progress that has been made in the increasingly automated generation of MLPs in the past two decades~\cite{P5399,P5842,P6058}, constructing MLPs is still not a black box task. For instance, it is important to be aware that MLPs can spectacularly fail as a consequence of their flexible functional form if they have not been carefully validated or if they are used beyond the range of structures they have been developed for. Therefore, the construction and the validation of MLPs should go hand in hand, and users need to know about the applicability and the limitations of a specific potential, which strongly depends on the underlying data set. Hence it is important to distinguish between the general capabilities of a MLP method and the performance of a specific parameterization for a given system. This subtle difference poses a significant challenge for an unbiased comparison of different methodologies. 
\\
In spite of the rapidly growing body of publications on MLPs and their applications, there are not many Tutorials in the literature \cite{P0834,P4444,P6346} addressing these issues. With this Tutorial we aim to fill this gap by discussing in detail all important aspects of training and validating MLPs. As a real-life example, we will use high-dimensional neural network potentials (HDNNP)~\cite{P1174} representing a common type of MLP. Nevertheless, most of the discussion equally applies to different classes of neural network potentials as well as MLPs employing other types of machine learning algorithms. The general procedures are illustrated using a simple model system consisting of a LiOH ion pair in a small box of water.
\\
The overall workflow for training a neural network potential is shown schematically in Fig.~\ref{fig:overview} and corresponds to the structure of this Tutorial. After a concise summary of the HDNNP methodology in Section~\ref{sec:hdnnp}, the generation of the initial data set is discussed in Section~\ref{sec:data}. Moreover, several decisions have to be made before the training process can be started, such as the selection of the features or descriptors and the settings of the neural networks, e.g. the architecture and activation functions. These preparations for the training are presented in Section~\ref{sec:initialization}. The training procedure, which is usually an iterative process that also involves the extension of the data set by active learning, is then explained in Sec.~\ref{sec:training} followed by a recipe for the validation of the potential in Sec.~\ref{sec:validation}. The Tutorial ends with some final remarks and conclusions in Sec.~\ref{sec:conclusion}. Some illustrative examples are given in the supplementary material.

\section{High-Dimensional Neural Network Potentials}\label{sec:hdnnp}

\subsection{Method \label{sec:method}}

\begin{figure*}[t]
    \centering
    \includegraphics[width=14cm]{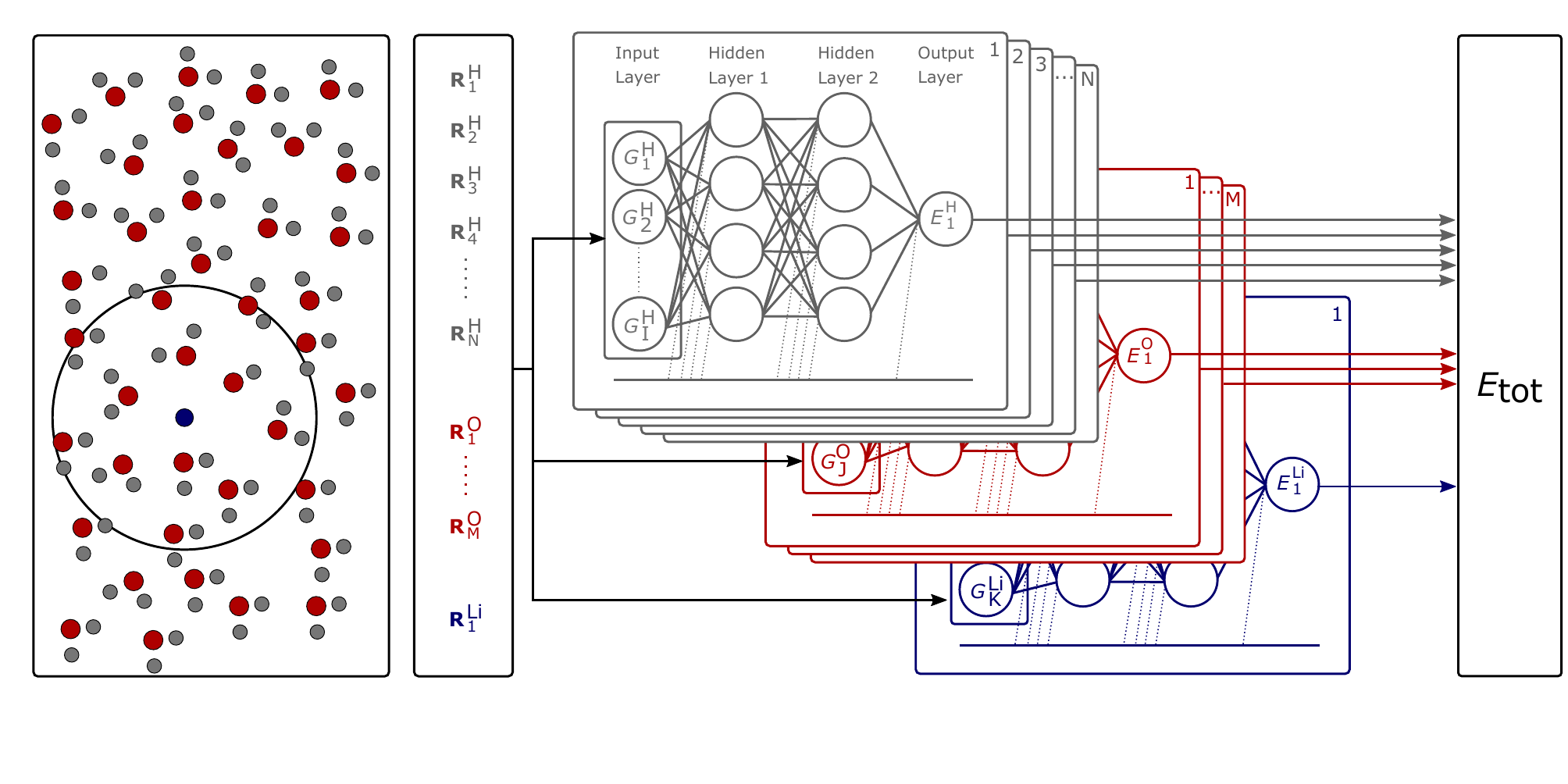}
    \caption{Schematic structure of a second-generation high-dimensional neural network potential (HDNNP)~\cite{P1174} for the example of a lithium hydroxide ion pair in water (lithium: dark blue; oxygen: red; hydrogen: grey). First, the Cartesian coordinate vectors $\mathbf{R}_i^{\alpha}$ of the atoms $i$ of element $\alpha$ are transformed to vectors of $j$ atom-centered symmetry functions (ACSF) $\mathbf{G}_j^{\alpha}$~\cite{P2882}, which describe the local atomic environments up to the cutoff radius shown as circle for the case of the lithium atom. For each atom the respective ACSF vector is then used as input for an atomic neural network (NN) predicting its atomic energy $E_i^{\alpha}$. Finally, the atomic energies are summed to obtain the total potential energy $E_{\mathrm{tot}}$ of the system (Eq.~\ref{eq:etot}). The sets of ACSFs, the architectures of the atomic NNs, and the weight parameters are the same for all atoms of a given element.}
    \label{fig:HDNNP}
\end{figure*}

A major drawback of early, first-generation MLPs has been the limitation to a small number of atoms, which prevented their use in simulations of complex molecular and condensed systems. A strategy to address large systems, which has been used in empirical potentials for a long time with great success is the decomposition of the total potential energy $E_{\mathrm{tot}}$ of the system into structure-dependent atomic energies $E_i$, e.g., in the famous Tersoff potential~\cite{P0820} and the embedded atom method~\cite{P0341}. 
\\
However, transferring this strategy to the realm of MLPs has been a frustrating task in the early days of MLP development. While the high dimensionality of machine learning algorithms is the reason for their superior accuracy, it also posed severe challenges for finding suitable coordinates considering the mandatory invariances of the potential energy with respect to translation, rotation and permutation, i.e., the order of chemically equivalent atoms of the same element. 
\\
The rotational and translational invariances can in principle be incorporated easily by using internal coordinates such as interatomic distances instead of the Cartesian positions. However, the output of machine learning algorithms such as neural networks (NN) depends on the order in which such internal coordinates are supplied in the vector of input values, which violates permutation invariance. Moreover, feed-forward NNs, which have almost exclusively been used in early MLPs, have a fixed dimensionality, preventing the construction of PESs for systems with variable numbers of atoms. 
\\
A solution to this problem has been found in 2007 with the introduction of atom-centered symmetry functions (ACSF)~\cite{P1174,P2882} as atomic environment descriptors fulfilling all required invariances. These enabled the development of high-dimensional neural network potentials~\cite{P1174} applicable to systems consisting of thousands of atoms using the total energy expression
\begin{eqnarray}
E_{\mathrm{tot}}=\sum_{i=1}^{N_{\mathrm{atoms}}}E_i  =\sum_{j=1}^{N_{\mathrm{elements}}}\sum_{i=1}^{N_{\mathrm{atoms}}^j} E_i^j  \quad . \label{eq:etot}
\end{eqnarray}
As the atomic environments determining the atomic energies are defined by a cutoff radius $R_{\mathrm{c}}$, HDNNPs of this form represent a second-generation potential. Today, a very large number of atomic environment descriptors~\cite{P3885,P6536,P5075,P5735,P5361} and many flavors of very accurate high-dimensional MLPs are available offering access to large-scale atomistic simulations of condensed systems.
\\
The structure of a HDNNP is shown in Fig.~\ref{fig:HDNNP} for the example of a LiOH ion pair in water. Starting from the Cartesian coordinate vectors $\mathbf{R}_i$ of the atoms first a transformation to the ACSF vectors $\mathbf{G}_i$ is carried out for each atom. In general, the number and definition of the ACSFs (cf. Sec.~\ref{sec:acsf}) can be different for each element. These vectors then enter the respective atomic feed-forward NNs (see Box 1), which are the central components of the HDNNP providing the atomic energies yielding $E_{\mathrm{tot}}$.
\\
For a given element, the architecture and weight parameters of the atomic NNs are constrained to be the same. Consequently, $N_{\mathrm{elements}}$ different feed-forward NNs need to be trained simultaneously, and in applications of the HDNNP each element-specific NN is evaluated once per each atom of the respective element. Apart from the element information and the atomic positions -- and if applicable the lattice vectors -- no further structural information about the system is required for second-generation MLPs. Moreover, in contrast to most traditional force field no further classification of atoms of a given element into atom types depending on the local bonding patterns is required.

%
\begin{mdframed}[style = mystyle,frametitle = {
Box 1: Feed-forward neural networks
}]
\label{box:NNP}
Multilayer feed-forward neural networks~\cite{B0001,B0002}, often also called ``deep neural networks'', are a type of artificial neural network whose functional form is inspired by the interaction of neurons in the brain \cite{P1713}. While initially they were used to develop mathematical models of these interactions, nowadays they have become a powerful technique in the field of machine learning to establish functional relations between input and target properties
~\cite{B0001, clarkScientificApplicationsNeural2014}.
\\
An example of a feed-forward neural network is shown in Fig.~\ref{fig:FFNN}. As its name implies, the flow of information is from left to right, i.e., from the input to the output layer along the  black lines connecting the neurons, or nodes, in the network, which are represented by the circles. The structural information is given by the vector $\mathbf{G}=\left\{G_i\right\}$ and is provided in the input layer. It is then passed via one or more hidden layers to the node(s) in the output layer. 
\\
Each hidden layer contains several neurons, which are connected to the neurons in the neighboring layers by weight parameters $a_{ij}^{kl}$ representing the connection strength between node $i$ in layer $k$ and node $j$ in layer $l=k+1$. In addition, each neuron $j$ in layer $l$, which may be a hidden or the output layer, is connected to a bias node by a bias weight $b^l_j$ acting as an adjustable offset.
\\
The numerical value $y_j^l$ of a node $j$ in layer $l$ is given by a linear combination of the values of all neurons $y_i^k$ in the previous layer weighted by the respective connecting weight parameters. Then, the bias weight is added and a non-linear activation function $f_j^l$ is applied yielding
\begin{eqnarray}
    y_j^l=f_j^l\left(b_j^l+\sum_{i=1}^{N_{\mathrm{nodes}}^k} a_{ij}^{kl} \cdot y_i^{k}\right) \quad .
\end{eqnarray}
Consequently, the output $E$ of the small feed-forward neural network in Fig.~\ref{fig:FFNN} is given by 
\begin{eqnarray}
    E&=&f_1^3\Bigg(b_1^3+\sum_{k=1}^4 a_{k 1}^{23} \cdot f_k^2\Bigg(b_k^2 \nonumber \\
    &+&\sum_{j=1}^4 a_{j k}^{12} \cdot f_j^1\left(b_j^1+\sum_{i=1}^3 a_{i j}^{01} \cdot G_i\right)\Bigg)\Bigg)\quad .
\end{eqnarray}

Activation functions are an important component of a neural network, since they introduce non-linearity to the model and thereby give feed-forward neural networks the ability to fit arbitrary non-linear functions\;\cite{kohonenIntroductionNeuralComputing, hornikMultilayerFeedforwardNetworks1989a} such as potential energy surfaces. Various functional forms can be used such as the hyperbolic tangent, the sigmoid function, and Gaussians. They have in common that they contain a non-linear region, and are continuous and differentiable, which is needed for the gradient-based optimization of the network and for the computation of atomic forces in MLPs. For the output layer usually a linear function is used to avoid restricting the output values of the feed-forward neural network. Alternatively, the target values can be (re)scaled.
\\
The flexibility of a network is determined by its architecture, i.e., the number of hidden layers and nodes per layer. The more layers and nodes are included, the larger are the fitting capabilites of the network, which often contain a few thousand weights. Typically, the feed-forward NNs in HDNNPs consist of two to three hidden layers including between 15 and 50 neurons each.

\begin{center}
    \includegraphics[width=7cm]{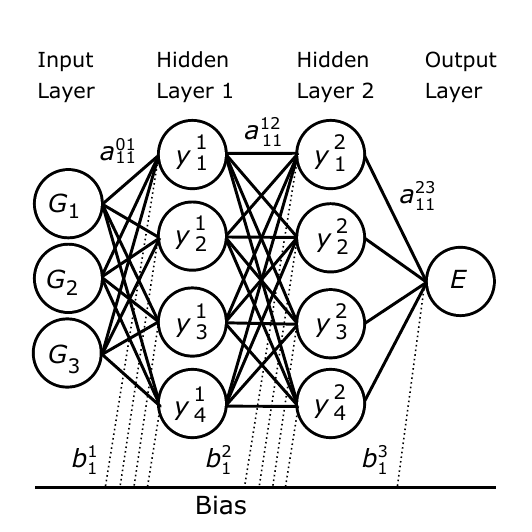}
      \captionof{figure}{Architecture of a feed-forward neural network. The output energy $E$ is a function of the neurons $\{G_i\}$ in the input layer. In between the input and the output layer the neurons are arranged in hidden layers and determine the functional flexibility of the neural network. The black lines connecting pairs of neurons and also dotted lines between the bias node and the neurons represent the fitting weight parameters of the network.}
     \label{fig:FFNN}
      \par
\end{center}
\end{mdframed}

\subsection{Atom-Centered Symmetry Functions \label{sec:acsf}}

\begin{figure*}[t]
    \centering
    \includegraphics[width=2\columnwidth]{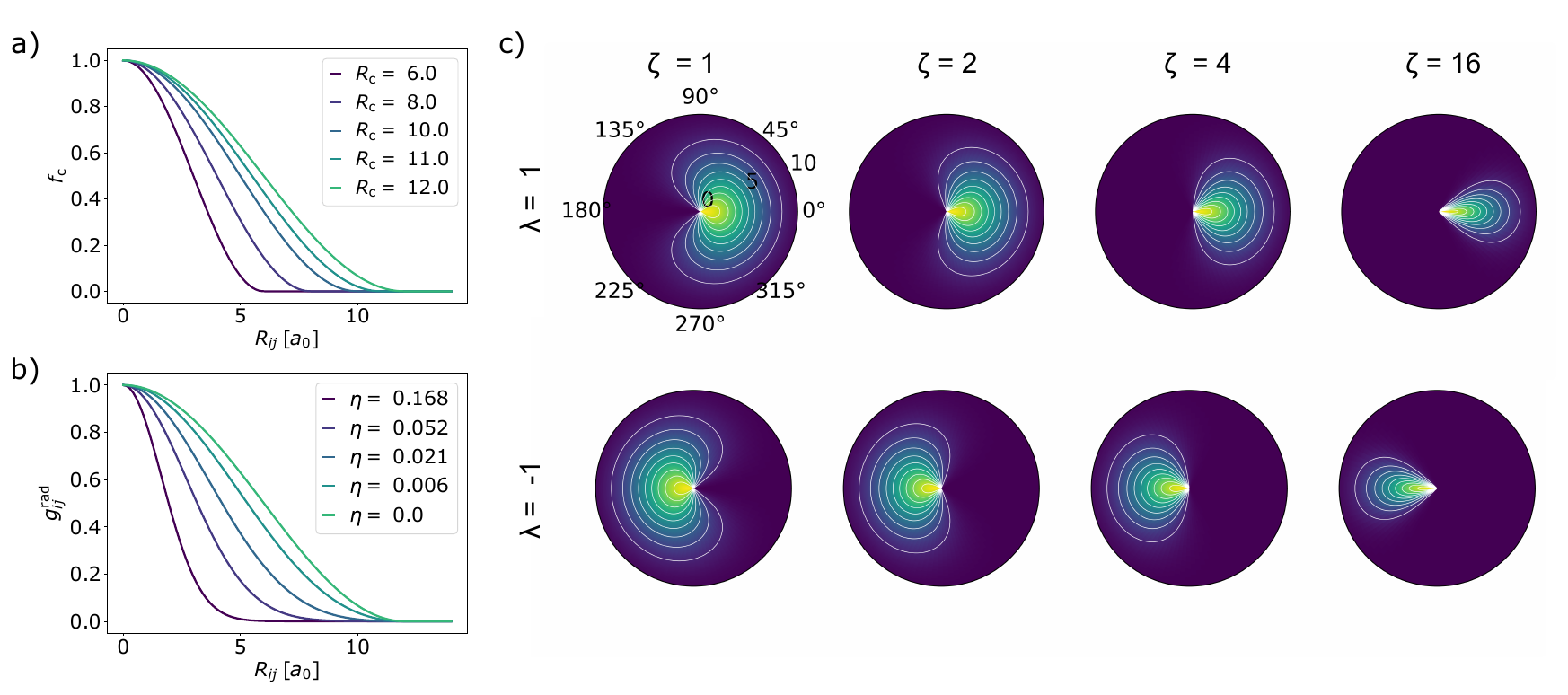}
    \caption{ Different types and components of atom-centered symmetry functions (ACSF). Panel a) shows the cutoff function $f_{\mathrm{c}}(R_{ij})$ (Eq.~\ref{eq:cutoff}) for different cutoff radii $R_\mathrm{c}$ (in $a_0$). Panel b) displays a term $g^{\mathrm{rad}}_{ij} =$ $e^{-\eta (R_{ij}-R_{\mathrm{s}})^2}$ $\cdot f_c(R_{ij})$ of the radial symmetry function  in Eq.~\ref{eq:rad_symfun} for different Gaussian exponents $\eta$ (in $a_0^{-2}$) with $R_{\mathrm{c}}=12\;a_0$ and $R_{\mathrm{s}}=0\;a_0$. In panel c) the dependence of the angular symmetry functions (Eq.~\ref{eq:ang_symfun}) on the angle $\theta_{ijk}$ and the distance $R_{ij}$ as radial coordinate is displayed for a cutoff of $R_{\mathrm{c}}=12\;a_0$ and the cases $\lambda=+1$ and $\lambda=-1$ with different $\zeta$ parameters.
   }
    \label{fig:symfunctions}
\end{figure*}

The availability of suitable descriptors for the atomic environments is a key for the construction of high-dimensional MLPs. Here, we will just briefly summarize two types of atom-centered symmetry functions, which are most commonly chosen when constructing HDNNPs, but in principle
many other types of descriptors could equally be used.
The spatial extension of the ACSFs as a function of distance $R_{ij}$ of atom $j$ from central atom $i$ is defined by a cutoff function such as the monotonously decreasing part of the cosine function,
\begin{align}
    f_{\mathrm{c}}\left(R_{i j}\right)=\left\{\begin{array}{cll}
0.5\cdot\left[\cos\left(\frac{\pi R_{ij}}{R_{\mathrm{c}}}\right)+1\right] & \text { for } & R_{i j} \leq R_{\mathrm{c}} \\
0.0 & \text { for } & R_{i j}>R_{\mathrm{c}}\,,
\end{array}\right.
\label{eq:cutoff}
\end{align}
which smoothly decays to zero in value and slope at $R_{\mathrm{c}}$ (cf. Fig.\;\ref{fig:symfunctions}a). There are two main classes of ACSFs termed radial and angular, which can be used to provide local structural fingerprints of the atomic environments.
\\
The most common radial ACSF has the form
\begin{align}
    G_i^{\mathrm{rad}}=\sum_{j=1}^{N_{\text {atoms}}\in R_{\mathrm{c}}} e^{-\eta\left(R_{i j}-R_{\mathrm{s}}\right)^2} \cdot f_{\mathrm{c}}\left(R_{i j}\right) 
    \label{eq:rad_symfun}
\end{align}
and consists of a set of Gaussian functions (see Fig.~\ref{fig:symfunctions}b) evaluated at the radial distances of all neighboring atoms inside the cutoff sphere. To ensure that the number of symmetry functions is independent of the number of neighbors, which is required for compatibility with the fixed dimensionality of the input vectors of the atomic NNs, the Gaussian functions are multiplied by the cutoff function and the sum of all terms is calculated to yield a single value, which can be interpreted as an effective coordination number within a range defined by the Gaussian width parameter $\eta$. The parameter $R_\mathrm{s}$ allows to shift the centers of the Gaussians away from central atom $i$ to increase the sensitivity of the function in specific spherical shells. To obtain a radial profile of the neighboring atoms, a set of radial symmetry functions is constructed employing element pair-specific sets of different $\eta$ values (see Sec.~\ref{sec:initialization}).
\\
Angular symmetry functions consider the angles $\theta_{i j k}$ of triplets of atoms centered between the connections $ij$ and $ik$ employing the cosine function such that the periodicity of the angle and in particular its mandatory symmetry with respect to angles of $0^{\circ}$ and $180^{\circ}$ are taken into account,
\begin{align}
    G_i^{\mathrm{ang}}=&2^{1-\zeta} \sum_{j \neq i}^{N_{\text {atoms}}\in R_{\mathrm{c}}} \sum_{k \neq i, j}^{N_{\text {atoms}}\in R_{\mathrm{c}}}[\left(1+\lambda \cdot \cos \theta_{i j k}\right)^\zeta\nonumber \\ &\cdot e^{-\eta\left(R_{i j}^2+R_{i k}^2+R_{j k}^2\right)}
    \cdot f_{\mathrm{c}}\left(R_{i j}\right) \cdot f_{\mathrm{c}}\left(R_{i k}\right)\cdot f_{\mathrm{c}}\left(R_{j k}\right)] \quad.
    \label{eq:ang_symfun}
\end{align}
There are two parameters, which can be varied to characterize the atomic environment. These are the values of $\zeta$ and $\lambda$. As can be seen in  Fig.\;\ref{fig:symfunctions}c, the purpose of $\lambda=+1$ or $\lambda=-1$ is to center the maxima of the cosine function at angles of $0^{\circ}$ or $180^{\circ}$, respectively, while a series of $\zeta$ values controls the angular resolution of the functions. Multiplying the angular terms by the three cutoff functions of the three pairwise distances ensures that only triples of close atoms enter the summation, while the Gaussian functions can be used to contract the spatial range to close neighbors, if desired. This can be useful since for a given angle the distance between neighbors $j$ and $k$ increases with $R_{ij}$ and $R_{jk}$ resulting in weaker interactions. 
\\
It is important to note that there is no need for a linear relation between the values of the ACSFs and the potential energy, since the atomic NNs are able to express very general functional forms. Therefore, only some mandatory properties of the ACSFs are relevant for the construction of HDNNPs (see Box 2). For each element in the system, its radial and angular functions are constructed for all element combinations present for atoms $j$ (radial functions) or $j$ and $k$ (angular functions), respectively, to reflect the different interactions between different chemical species. Typically, between 10 and 200 ACSFs are used per atom depending on the complexity of the system. Although ACSFs consist of atomic distances and angles, they simultaneously depend on all atomic positions inside the cutoff sphere. Hence, they are formally many-body functions. Still, it has been shown that in rare situations descriptors consisting of two- and three-body terms may not be unique in particular for low-dimensional systems~\cite{P5765}. A more detailed discussion and further types of ACSFs can be found in Ref.~\citenum{P2882}.

\begin{mdframed}[style = mystyle,frametitle = {Box 2: Properties of ACSFs}]
To be suitable descriptors for the atomic environments, atom-centered symmetry functions must fulfill several requirements.:
\begin{itemize}
    \item They need to describe the structural details of the atomic environments.
    \item Their values must be invariant with respect to rotation, translation and permutation.
    \item They have to decay smoothly to zero in value and slope at the cutoff radius to restrict the atomic interactions to the environments inside the cutoff spheres.
    \item They need to be continuous and differentiable for the gradient-based training of HDNNPs and the calculation of analytic forces.
    \item The number of ACSFs in the atomic NN input vectors must be independent of the number of neighbors inside the cutoff sphere, because this number can be different for each atom and can change in MD simulations.
\end{itemize}
\end{mdframed}

\subsection{Long-Range Electrostatic Interactions \label{sec:methodelectrostatic}}

So far, we have discussed the structure of second-generation HDNNPs and ACSFs as descriptors for the training of environment-dependent atomic energies. The main assumption of such second-generation potentials is the locality of a major part of the atomic interactions, which, if valid, can be expressed to a good approximation in terms of local atomic energies.  For clarity it should be noted that even second-generation MLPs contain all types of interactions, including electrostatics and dispersion, between atoms within the cutoff radius, since Eq.~\ref{eq:etot} does not distinguish between different types of bonding. However, while this ansatz works surprisingly well for many systems, in some cases it is necessary to explicitly include long-range interactions such as electrostatics without truncation~\cite{P5975}.
\\
The inclusion of long-range electrostatic interations is possible using, e.g., an Ewald sum~\cite{P0238} employing atomic charges determined by ML. The extended total energy expression is then given by
\begin{eqnarray}
E_{\mathrm{tot}}=E_{\mathrm{elec}}+E_{\mathrm{short}}=E(\{Q_i\},\{\mathbf{R}_i\})+\sum_{i=1}^{N_{\mathrm{atoms}}}E_i(\{\mathbf{R}_i\})     \label{eq:eelec}
\end{eqnarray}
as a sum of a short-range and a long-range part.
\\
There are different options to obtain the required charges. If the atomic charges are essentially local, i.e., they only depend on the close chemical environment, they can be learned as atomic properties using ACSF descriptors and a second set of atomic neural networks, e.g., in third-generation HDNNPs~\cite{P3132,P2962}. If, however, the charges depend on distant structural features and are influenced by long-range charge transfer, such as in aromatic systems or molecules containing conjugated $\pi$-bonds, fourth-generation potentials may be required~\cite{P5932}. In this case, the charges are determined indirectly via environment-dependent atomic electronegativities in combination with a charge equilibration step or a self-consistent redistribution of the charges. 
\\
Most of the aspects of training neural network potentials are not much affected by the decision if a second-, third-, or fourth-generation HDNNP is trained, since the procedures for the selection of the data, the iterative training and the validation are essentially the same. Therefore, in this Tutorial we will focus on the example of a second-generation HDNNP, while some comments on the particularities of training HDNNPs including electrostatic interactions are given in Section~\ref{sec:beyondshortranged}. Further details on third- and fourth-generation HDNNPs can be found in Refs.~\citenum{P3132,P2962,P5932,P5977,P6018}.

\section{Data Set Generation}\label{sec:data}

\subsection{Reference Electronic Structure Calculations \label{sec:reference}}

\begin{figure*}[ht]
    \centering
    \includegraphics[width=14cm]{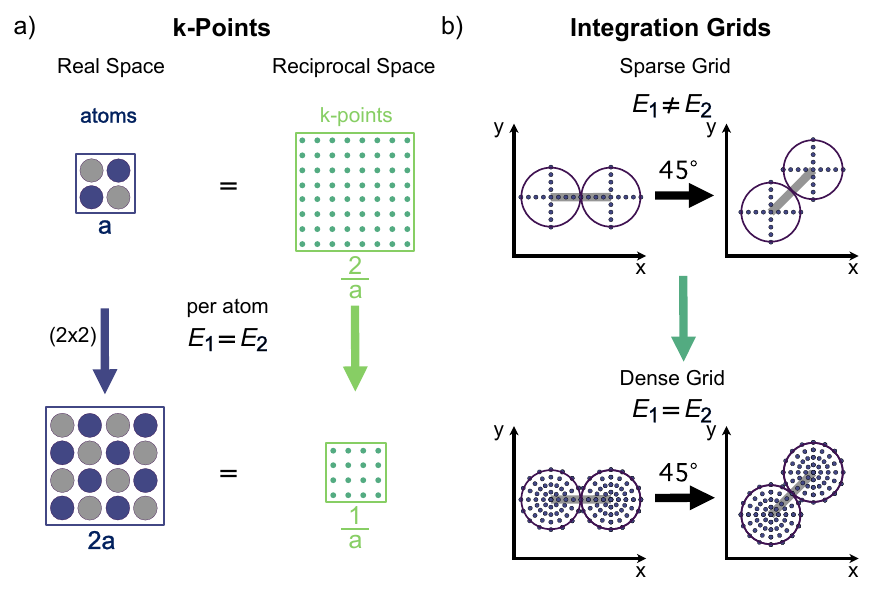}
    \caption{Numerical accuracy of integration grids in electronic structure calculations. Panel a) shows the selection of consistent k-point grids in periodic DFT calculations when changing the cell size. For supercells it is sometimes possible to choose exactly equivalent k-point meshes in reciprocal space, e.g., when the lattice parameter $a$ is doubled and the number of k-points in the corresponding direction can be devided by two. For the combination of arbitrary systems of different size in the training set the k-point density should be kept constant. Panel b) illustrates the role of numerical integration grids for the case of a diatomic molecule, which might, e.g., be atom-centered grids used for the representation of numerical atomic orbitals or electron densities. For a sparse grid, the energy of the system does not show the required rotational invariance (e.g. in representing the grey region), while for a dense grid this invariance is to a good accuracy achieved. The same effect can be observed for regular three-dimensional grids used, e.g., in Fourier transformations.}
    \label{fig:grids}
\end{figure*}

Before a MLP can be trained, first a reference electronic structure method has to be chosen. When constructing potentials for condensed systems, often DFT calculations are used. These  can be carried out routinely for systems containing a few hundred atoms with and without periodic boundary conditions even if thousands of data points are required, which is typical for MLP training. However, also coupled cluster and other wave function-based methods are increasingly popular~\cite{P6360,P6362}, since any limitation in the accuracy of the training data will become an intrinsic property of the MLP, which thus cannot provide more accurate results than the underlying reference method. Still, due to the high computational costs, often a compromise between accuracy and computational efficiency has to be found when choosing the reference method. It is thus important to note that the choice of the reference method determines the reliability of the final MLP, i.e., the agreement with experimental observables, and the only aim of MLP construction is to reproduce this method with all its intrinsic properties.
\\
Ideally, the reference method should not only provide accurate energies but also atomic forces, which contain a wealth of information about the local shape of the PES. Including forces in the training, which is good practice in almost any modern MLP~\cite{P0834,P2630,P3114,P6065,P5795}, offers several advantages. First, the amount of information that can be extracted from expensive electronic structure calculations is dramatically increased, since each single point calculation can only provide one energy value but $3N_{\mathrm{atoms}}$ force components. Moreover, as has been shown recently for the case of metal-organic frameworks~\cite{P5537}, MLPs can be subject to arbitrary internal energy shifts between the individual atoms.
These internal energy shifts, which can strongly reduce the transferability of MLPs, are invisible when monitoring the total energy errors in the training process, because there are many ways to distribute the total energy in the system for a given total energy. Forces, which can provide local information about the gradients of the PES at the atomic positions, can contribute to overcome these problems.
\\
The generation of the reference data set is often the computational bottleneck in the construction of MLPs as they usually consist of tens of thousands of data points. This effort can be reduced in two ways: by carefully choosing and thus decreasing the number of structures, which is a very active field of research (see Sec.~\ref{sec:activelearning}), and by reducing the costs of each calculation by optimizing the settings of the employed electronic structure code. 
\\
The numerical convergence of the electronic structure calculations is a crucial point, since any numerical noise in the data represents a limit to the minimum error that can be reached in the training process. Particularly problematic is the choice of k-points if periodic systems of different size or symmetry are combined in a single training set, because underconverged k-point grids can have a significant impact on the accuracy of energies and forces, which finally results in inconsistent data in the reference set (see Fig.~\ref{fig:grids}a). Attention should also be paid to the density of Furthermore numerical integration grids, which usually have a fixed orientation in space~\cite{P0141,P2577}. If too sparse grids are used, data sets with different molecular orientations might contain similar structures with contradictory energies and forces (see Fig.~\ref{fig:grids}b). While this is straightforward to detect for small gas phase molecules, in condensed systems such as liquid water, which contain many similar bonds in multiple different directions, such inaccuracies might be extremely difficult to detect resulting in poor potentials. In general, a numerical convergence to about 1 meV/atom for energy differences and 0.05--0.10 eV/\AA{} for the forces should be considered as a minimum requirement.
\\
Finally, there are other properties next to energies and forces, which might be needed for the construction MLPs and should be computed and stored. Apart from atomic partial charges for MLPs containing electrostatic interactions, this might be information about the spin for magnetic systems~\cite{P5867,P6057,P6509} or Hirshfeld volumes for dispersion interactions~\cite{P6537}. Sometimes also the stress tensor is used for training~\cite{P5828}.

\subsection{System Size and Dimensionality}

\begin{figure*}[ht]
    \centering
    \includegraphics[width=15cm]{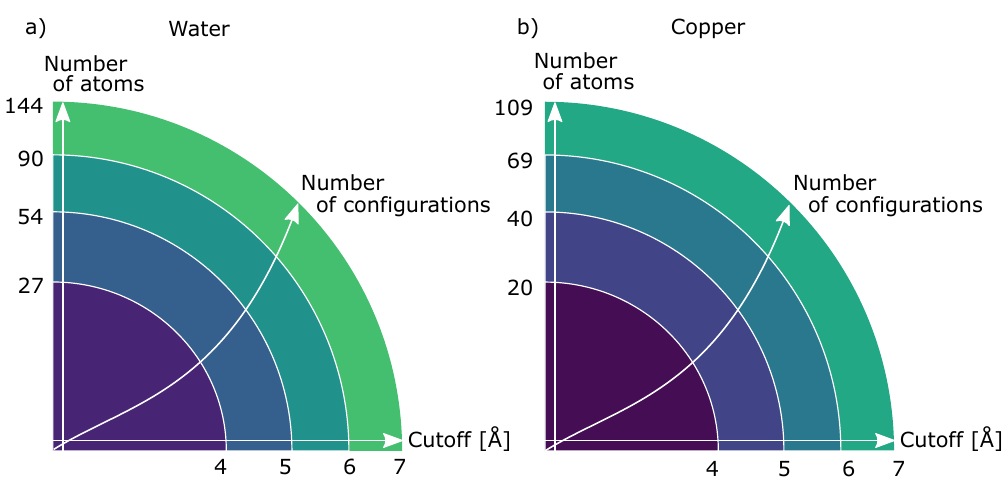}
    \caption{Structural complexity of atomic environments. Panels a) and b) show typical numbers of atoms inside the cutoff spheres for different cutoff radii $R_{\mathrm{c}}$ for bulk liquid water and fcc bulk copper at ambient conditions. The number of possible atomic configurations inside the cutoff spheres increases strongly with the effective dimensionality of the atomic environments. }
    \label{fig:atomnumbers}
\end{figure*}

Once the decision concerning the electronic structure method has been made and converged settings have been determined, the required structure types have to be selected. This includes many aspects, such as the system size, the chemical composition and the structural diversity, because MLPs have limited extrapolation capabilities beyond the range of structures included in the training set. Unfortunately, with the exception of smallest molecules in vacuum, mapping atomic configurations on a regular grid is not a feasible option due to the exponential growth in the amount of possible structures with increasing number of atoms.
\\
The construction of the total energy as a sum of environment-dependent atomic energies in second-generation MLPs corresponds to an effective reduction of the dimensionality for large condensed systems since only the positions of the atoms inside the cutoff spheres need to be learned while the total energy of the system still explicitly depends on all atomic positions. Fig.~\ref{fig:atomnumbers} shows the number of neighboring atoms included in the local environment for cutoff radii between 4 and 7~\AA{} for the examples of liquid water and fcc bulk copper. As can be seen, typical environments contain between 50 and 150 atoms. These atoms define the configuration space that should be mapped for the generation of the MLP, which requires a system size of at least the extension of the cutoff sphere. In case of periodic systems, the smallest cell diameter should be larger than the diameter of the cutoff sphere to avoid sampling only a subspace of the possible atomic configurations due to an artificial periodicity of the system. Thus, training structures for condensed systems often reach a size between 150 and 250 atoms. Still, it is possible to combine systems of different size in the training set to benefit from faster calculations for smaller systems, as long as consistent settings are used (see Sec.~\ref{sec:reference}). 
We note here that also MLPs of the third and fourth generation follow the same principle of environment-dependent properties such that also for these generations in principle it is not necessary to significantly increase the size of the reference structures. However, the choice of the cutoff is a critical decision and will be discussed in Sec.~\ref{subsec:Descriptor}.
\\
Even if the effective dimensionality is reduced due to the cutoff sphere, a comprehensive sampling is still out of reach and systematic strategies to select the most important data points are needed (see Sec.~\ref{sec:activelearning}). Unphysical atomic configurations, such as too close atomic encounters or chemically unreasonable bonding patterns should be avoided, as they would have very high potential energies and would be irrelevant in atomistic simulations. Still, any moderate and thus chemically still relevant increase in temperature enlarges the size of accessible configuration space and makes the mapping more demanding.
For some condensed and molecular systems it is also possible to reduce the complexity of the reference systems by making use of molecular fragments, which can be extracted from the system~\cite{P4585,P5537,P6065,P6527}.

\subsection{Generation of an Initial Data Set \label{sec:initialdataset}}

Before a first MLP can be trained, an initial data set needs to be available, which -- depending on the complexity of the system -- might contain a few hundred up to a few thousand structures. Ideally, the atomic configurations should be chemically reasonable and as close as possible to the geometries visited in the intended applications of the potential. They should therefore be sampled at the corresponding conditions, such as temperature, pressure, and chemical composition.
\\
A straightforward although expensive way of generating reasonable initial data sets is the use of ab initio MD. Here, often less tight settings are used, such as $\Gamma$-point sampling of k-space only. Therefore, energies and forces from ab initio MD trajectories are often not used directly but provide geometries, which are then recomputed with a higher degree of convergence or even more demanding electronic structure methods. This also allows to avoid the computation of structurally correlated geometries such as configurations visited in subsequent MD steps, which only provide a marginal amount of new information, by selecting structures in specific intervals. 
\\
A significant advantage of ab initio MD is the generation of reasonable structures already right in the beginning of the data set construction. Still, it should be taken into account that potential energy wells are more frequently sampled than repulsive high-energy regions or transition states. The representation of repulsive walls can be improved by running simulations at slightly elevated temperatures and pressures, which increase the probability of closer atomic encounters. Transition states can be included using, e.g.,   metadynamics~\cite{P0545}. 
\\
If available for the system of interest, first structures for electronic structure calculations can also be extracted from conventional force field trajectories. It should be noted, however, that different equilibrium geometries may result in structural biases, which -- to a much smaller extent -- can also be present in ab initio MD simulations when using less converged settings or different functionals.
\\
There are several alternative approaches to generate preliminary data sets. In particular in the field of materials science it is common to start from the known crystal structures of a material and to introduce step-by-step random atomic displacements of increasing amplitude to sample the local minimum wells. In molecular systems, also normal mode sampling is frequently used~\cite{P4945,P6538}. Furthermore, randomly placed atoms and molecules can be used in combination with suitable interatomic distance constraints. Finally, also the increasing availability of repositories and databases might be an interesting source of initial data sets~\cite{P6539,P6540}. However, the initial data set represents only a first step, and its extension by further consistent data is crucial, which makes the reuse of repository data difficult, as the exact input settings need to be known and the same version of the same electronic structure code has to be available (see Sec.~\ref{sec:activelearning}).

\begin{mdframed}[style = mystyle,frametitle = {Case study: Construction of a first data set for the LiOH-water system}]
For the DFT calculations of the lithium hydroxide system the Car-Parrinello Projector Augmented Wave (CP-PAW) code (version 28-09-2016) was employed \cite{CPPAW} using the PBE0r local hybrid functional~\cite{P5071} employing the settings described in Ref.~\citenum{P5861}. CP-PAW implements the projector augmented-wave (PAW) method \cite{P0297}, in which augmentations of the 1s orbital of H, the 2s and 2p orbitals of Li, and the 2p, 2p, and 3d orbitals of O were used. The auxiliary wave functions were set up as node-less partial waves. Their matching radii for all orbitals are 0.7 times the covalent radii, which are $0.32\;\mathrm{\AA}$ for H, $1.23\;\mathrm{\AA}$ for Li, and $0.73\;\mathrm{\AA}$ for O. The plane wave cutoffs of the auxiliary wave functions and the auxiliary density have been set to $35\; E_{\mathrm{H}}$ and $140\;E_{\mathrm{H}}$, respectively. In the PBE0r tight binding orbitals used for the calculation of the Hartree-Fock exchange, contributions of the 1s orbital of H, the 2s orbital of Li, and the 2s and 2p orbitals of O were incorporated. 
\\
For cubic cells with a lattice constant of $8.2\;\mathrm{\AA}$ a $2 \times 2 \times 2$ k-point grid was chosen. k-point grids of similar k-point density were selected for other cell sizes. Wave function and geometry optimizations were carried using the Car-Parrinello ab initio molecular dynamics method \cite{P0433} including a friction term. The total energy was converged to out up to a numerical convergence of $10^{-5}\; E_{\mathrm{H}}$.
The D3 dispersion correction~\cite{P3112} was applied by the DFTD3 software (version 14-06-2016), which uses Becke-Johnson damping.
\\\\
The periodic system contains one lithium ion, one hydroxide ion and between 19 to 23 solvating water molecules. We note that these systems are too small to provide a realistic description of general LiOH solutions and have been chosen for demonstration purposes in this Tutorial only. To construct an initial data set, at first the lithium ion, the hydroxide ion, and the water molecules were placed randomly in small boxes. The dimensions of the box were chosen to yield densities in the range between 0.97 and 1.02 kg\,dm$^{-3}$. Structures were checked and only accepted if atom-atom distances were larger than element-pair specific minimum distance thresholds. These thresholds are 1.2 \AA{} for H to H, 0.85 \AA{} for H to O and H to Li, and 2.3 \AA{} for O to O. \\ 

Initially, the structures have been optimized. Then, ab initio MD simulations with different simulation temperatures ranging from 100\;K to 500\;K were set up. After 5000 equilibration time steps every 1000 time steps the configurations were saved. In this way an initial data set of 630 reference structures was created. To this first data set an existing set of pure water structures reported in Ref.~\citenum{P6141} was added.\\\\

\end{mdframed}

\section{Preparations for the Training}\label{sec:initialization} 

\subsection{Preparing the Atomic Neural Networks \label{sec:NNpreparation}}

With an initial data set at hand, several further choices have to be made before the training process can begin. First of all, the architectures of the individual element-specific atomic neural networks have to be chosen. Often, several architectures differing in size are used for training, because a priori it is not clear, which NN architecture, i.e., which number of hidden layers and neurons per hidden layer, will be best suited for representing the data set. Moreover, the non-linear activation functions have to be selected. This process will then generate a set of different HDNNPs, which are also needed for the iterative improvement of the data set by active learning (see Sec.~\ref{sec:activelearning}). 
\\
In principle, for each element a different NN architecture could be defined, but for simplicity it is common to use the same architecture for all elements, which typically consists of two -- sometimes three -- hidden layers, each of which contains between 15 and 50 neurons per layer. Therefore, compared to other applications in data science, the atomic NNs in HDNNPs are relatively small and can be quickly evaluated. The atomic NNs are usually standard, i.e., fully connected, feed-forward NNs, but also modifications such as direct links from the input to the output layer or omitted connections are possible.
\\
When choosing the architecture, care should be taken that the number of weight and bias parameters defined in this way is smaller than the information content of the training set. Although the atomic NNs typically contain a few thousand parameters for each element, this is usually not a problem, if also the wealth of information included in the forces is used for training. 
\\
Another point to be considered regarding the size of the atomic NNs is the increasing flexibility as a consequence of the growing number of parameters in larger atomic NNs. If the NNs are too small, the flexibility is insufficient and the HDNNP will not be able to resolve all the details of the PES even if this information is present in the data set, which is called ``underfitting''. If the NNs are too large, ``overfitting'' will occur resulting in a poor prediction quality for structures not included in the training set. A more detailed description of the detection of overfitting and the selection of NN architectures is given in Sec.~\ref{sec:optimization}.

\subsection{Atom-Centered Symmetry Functions}\label{subsec:Descriptor}

Next, the atom-centered symmetry functions have to be chosen. As explained in section\;\ref{sec:method} the purpose of the descriptors is to provide a structural fingerprint of the environments inside the cutoff spheres while being invariant with respect to permutational, translational and rotational symmetry. In contrast to the atomic NN architectures, the ACSFs are usually not the same for each element, because the typical bond distances between atoms of different elements depend on the specific chemical species.
\\
First, the cutoff radius defining the atomic environments has to be selected. It represents a convergence parameter that has to be increased until all important atomic interactions are included. For a too small cutoff, atoms outside the cutoff sphere are still significantly interacting with the central atoms, which is equivalent to noise in the data limiting the accuracy that can be reached, since no information about the positions of the atoms outside the cutoff sphere is available in the ACSFs. If the cutoff is too large, the construction of the MLP becomes more demanding, since a large configuration space in the atomic environments has to be mapped by the reference electronic structure calculations. Consequently, for large cutoffs the effort to be invested in the reference calculations increases with respect to both, the number of structures and the size of the reference systems. Moreover, the detailed representation of the atomic positions inside the cutoff spheres requires a larger number of descriptors, which increases the computational costs of the HDNNP energy and force prediction. 
\\
Several procedures to determine the size of the cutoff radius, which is required for a certain level of convergence of the forces acting on the central atom, have been suggested in the literature. Since atomic interactions beyond the cutoff introduce noise in the forces, the variance of the forces when freezing the atoms inside and moving atoms outside the cutoff sphere can be used for uncertainty quantification~\cite{P2630,P4958}. Only if this uncertainty is small, the cutoff is large enough. Convergence tests of the forces as a function of the environment have also been reported in Ref.~\citenum{P6065}. An alternative approach, in which the individual interaction strength with each neighboring atom can be determined, is based on the Hessian~\cite{P6312}, which provides the derivative of the atomic forces with respect to all atomic positions in the system. This method is also applicable to crystalline environments since small net forces may also be a consequence of compensating interactions in symmetric environments. 
\\
Interestingly, as discussed in Refs.~\citenum{P5128} and \citenum{P6527}, the analytic forces of second-generation MLPs have twice the spatial range of the cutoff spheres defining the atomic energy contributions. Since the forces in MLPs can be constructed as the analytic derivatives of the atomic energies (Eq.~\ref{eq:etot}), it is therefore possible to use reference systems, which are large enough to learn the atomic energies but too small to provide forces of a condensed phase environment. This has been demonstrated recently for the benchmark case of metal-organic frameworks~\cite{P6527} and can, if carefully tested, reduce the costs of the electronic structure calculations for the training set in that only information from small systems is used, while the obtained potentials can predict correct forces beyond the training range.
\\
Once the cutoff has been determined, which is typically between 5 and 8 \AA{}, the next step is the selection of the parameters of the ACSFs. Overall, there are two main strategies for the determination of these parameters. The first strategy aims for a balanced and systematic description of space in the atomic environments, similar to basis sets in electronic structure calculations. The second strategy is data-centered and aims to identify the best set of ACSFs for a given data set. This is motivated by the typically very inhomogenous distribution of atoms in the cutoff spheres, which is governed by chemical bonds and thus does not require the capability to describe all imaginable geometries. 
\\ 
Since the reference data sets employed in the construction of the potential are not static but are continuously extended by active learning (see Sec.~\ref{sec:activelearning}), often a combination of both strategies is used. In early stages unbiased and very general ``default'' sets of ACSFs (see Box 3) are generated following the first strategy, while for large and converged data sets finally the optimum set of ACSFs is used according to the second strategy to yield the smallest number of descriptors providing the desired accuracy. These can be selected in different ways, such as optimizing the ACSF parameters by genetic algorithms~\cite{P5292}, principal component analysis~\cite{casierUsingPrincipalComponent2020}, or selecting functions from large pools of candidates based on CUR decomposition or the Pearson correlation in an automatic way~\cite{P5398}.
\\
Different ACSFs can have very different ranges of values. It is therefore common to rescale the ACSF values before using them as input for the atomic NNs~\cite{P0830}. This can be done by subtracting the mean value and adjusting the standard deviation of the symmetry functions of the reference data set. This procedure has to be repeated whenever new structures are included in the training set, because the composition of symmetry function values will change with the data set. In particular in early stages of the training set constructions, there might also be individual ACSFs with a very small range of values. Such symmetry functions should be preliminarily eliminated from the set of functions, because a narrow range of values prevents proper scaling and can result in unstable training by learning numerical noise. Once the structural diversity in the data set increases resulting in an extended range of values, these ACSFs should be included again.
\\
For all these reasons, the set of ACSFs is not at all fixed in the training process but has to adapt to the increasing data set size for optimum performance, which can in principle be done in an automatic way. The final dimensionality of the ACSF vector is usually smaller than the formal dimensionality of the configuration space in the atomic cutoff spheres but strongly depends on the chemical composition of the system. Since the radial and angular symmetry functions contain all combinations of elements in the system, the number of descriptors strongly increases with the number of elements in the system.

\begin{mdframed}[style = mystyle,frametitle = {Box 3: Constructing a default set of ACSFs}]
In early stages of HDNNP construction, often a default set of ACSFs is used to provide a balanced description of all possible configurations in the atomic environments.\\
\textit{Procedure for radial symmetry functions:}
\begin{itemize}
    \item For each element combination in the data set determine the minimum interatomic distance $R_{\mathrm{min}}$.
    \item For each element combination set $\eta_{\mathrm{max}}=0$~a$_0^{-2}$ to define the radial function with the largest spatial extension.
    \item Determine for each element combination the value of $\eta_{\mathrm{min}}$ such that the inflection point of the term $g_{ij}=e^{-\eta_{\mathrm{min}}R_{ij}^2}\cdot f_{\mathrm{c}}(R_{ij})$ is located at $R_{\mathrm{min}}$. If this inflection point is close to $R_{\mathrm{c}}$ the cutoff should be increased.
    \item Select further $\eta$ values between $\eta_{\mathrm{max}}$ and $\eta_{\mathrm{min}}$ for each element combination to obtain overall 5-6 radial functions with equidistant inflection points to yield a balanced radial resolution.
\end{itemize}
Since by construction the minimum interatomic distance has been taken into account, all functions should have a reasonable range of values enabling the scaling of the ACSFs. Radial functions of element combinations, which are not present in the system, are left out.\\
\\
\textit{Procedure for angular symmetry functions}
\begin{itemize}
    \item Identify all possible element triplets.
    \item Set the exponent $\eta=0$ a$_0^{-2}$ in Eq.~\ref{eq:ang_symfun} to generate angular functions with maximum spatial extension.
    \item For each triplet generate angular functions with $\zeta = 1,2,4,16$ for an approximately equidistant angular resolution.
    \item Combine each function with $\lambda =1,-1$ to yield in total eight angular functions per element triple.
    \item Check for each angular symmetry function the range of values and eliminate those functions with a too small range.
\end{itemize}
Optionally, a second set of contracted angular symmetry functions can be constructed using a larger value of $\eta$ to increase the angular sensitivity close to the central atom.
\vspace{1cm}
\end{mdframed}

\subsection{Initial Weight Parameters and Data Preparation \label{sec:weightsanddata}}

\begin{figure}[!ht]
    \centering
    \includegraphics[scale=0.60]{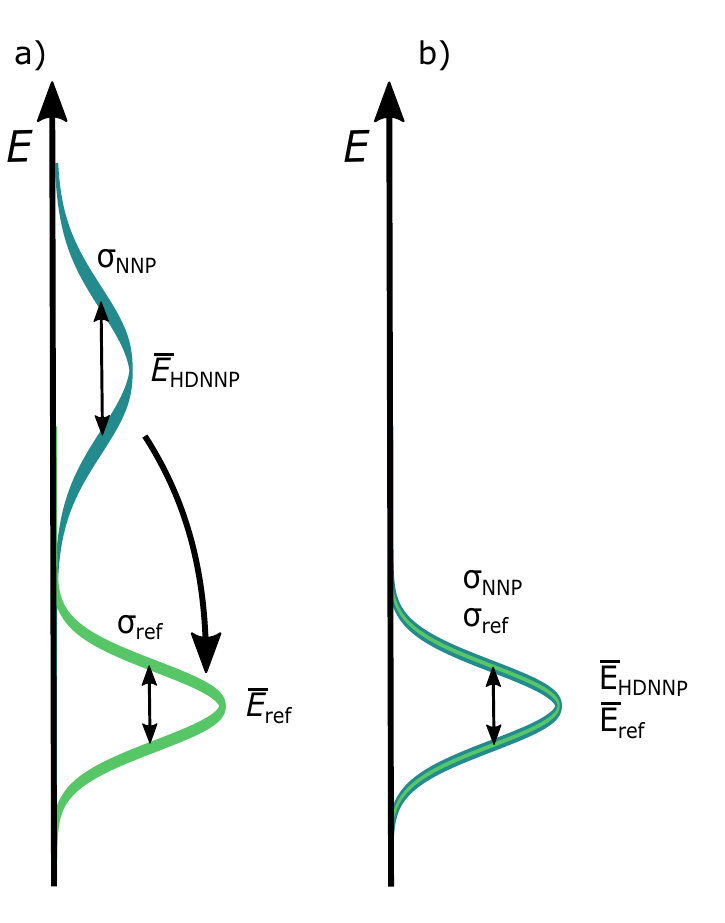}
    \caption{Preconditioning of the neural network weights. 
    In a first step, the mean energies $\bar{E}$ and the standard deviations of the energies $\sigma$ of the training data are determined for the HDNNP and the reference method (a). Then, the weight parameters are adjusted to match the respective values of the reference method to minimize the root mean squared error at the beginning of the training process (b).}
    \label{fig:preconditioning}
\end{figure}

Finally, while the number of weight parameters is defined by the architectures of the atomic NNs, a set of initial weight parameters has to be chosen as starting point for the HDNNP training. Apart from random numbers, e.g. in the interval $[-1,1]$, with normal or Gaussian distribution, several different methods for the selection of the starting weights have been proposed, such as the Nguyen Widrow scheme~\cite{P5742}, in which weights are chosen such that each hidden node approximates a part of the target function, or the Xavier method~\cite{glorotUnderstandingDifficultyTraining2010}, which aims to avoid values in the saturating region of the activation functions and is therefore especially suited for funnel-like architectures.
\\
However, the choice of the intial weights a priori does not ensure that the predicted energies and forces are close to the respective target values. This can be achieved by an additional preconditioning step before the training (see Fig.~\ref{fig:preconditioning}). If the initial weights are chosen randomly, the predicted mean energies and their standard deviations often differ significantly from the respective values of the reference data giving rise to a large error at the beginning of the training process. This error can be reduced by adjusting the initial weights such that the deviations between the mean energies of the HDNNP and the reference data as well as the differences in their standard deviations disappear. For this purpose, the mean HDNNP energy can be shifted by adjusting the bias weights of the output neuron, while the standard deviation can be controlled by additionally modifying the connecting weights between the neurons of the last hidden layer and the output layer.

Further, some preprocessing is performed for the reference data. Generally, splitting into a training and a test set is done (see Sec.~\ref{sec:optimization}), and also a third validation set is often used. The difference between the test and the validation set is that the test set is used to identify the potential with the minimum error for unknown structures. Since, however, the test set is thus involved in the selection of the potential, the latter is not completely independent of the test set. For this reason, the validation set might provide a more unbiased estimate for the generalization properties of the potential. However, as discussed in Sec.~\ref{sec:validation}, this splitting is not a reliable measure for the transferability of a potential since it is based on the assumption that the available reference set is covering the entire configuration space realiably, which cannot be taken for granted.
\\
Another preparation of the data set that is used if the range of output values is restricted by a hyperbolic tangent or sigmoid activation function in the output neuron is the scaling of the target energies to a unit interval, which in the application of the potential then has to be reversed. Moreover, depending on the choice of the reference method, the target total energies can have very large values, which is numerically inconvenient. Therefore, often the energies of the free atoms of all elements in vacuum are removed from the data set such that instead numerically much smaller binding energies are used in the training. The resulting offset can then be corrected by adding the corresponding free atom energies when applying the potential to yield total energies, which are numerically consistent with the reference method.

\begin{mdframed}
[style = mystyle,frametitle = {Case study: Symmetry functions values, architecture and weights initialization of the LiOH-water system}]
The symmetry function parameters of the LiOH-water system have been chosen following the recipe for generating a default set of ACSFs given in Box 3. As the values of the parameters $\eta$ of the radial symmetry functions depend on the minimum distances between the atoms of the respective element pairs in the reference data set, the set of symmetry function values has been adjusted throughout the construction of the HDNNP. In Table\;\ref{TAB:symfun_values} the final $\eta$ parameters of the converged potential are given, while the cutoff has been set to $R_{\mathbf{c}}=6$~\AA{} and a shift of $R_{\mathbf{s}}=0$~\AA{} has been used. The parameters of the angular symmetry functions have been selected as described in Box 3. In addition to the set of angular symmetry functions which has been derived following this procedure a second set of angular symmetry functions with the same values for $\xi$ and $\lambda$ but a value of $\eta = 0.025$ a$_0^{-2}$ has been employed to better describe the interaction of close atoms. Since there is only one lithium ion in the system and since the lithium ions in the periodic images are outside the cutoff radius, radial and angular ACSFs involving Li-Li interactions have been removed. 
\begin{center}
\captionsetup{width=0.95\linewidth}
 \captionof{table}{$\eta$ values of the radial symmetry functions of the HDNNP for LiOH in water.}
\label{tab_eta_rad_mn3_2}
\vspace{0.25cm}
\begin{tabular}{cc}
\toprule
Element pair & $\eta$ [a$_0^{-2}$]\\
\midrule
H-H&0.000, 0.006, 0.016, 0.038, 0.099\\
H-O&0.000, 0.007, 0.019, 0.051, 0.166\\
H-Li&0.000, 0.005, 0.012, 0.025, 0.052\\
O-O&0.000, 0.004, 0.008, 0.015, 0.027\\
O-Li&0.000, 0.005, 0.012, 0.024, 0.051\\
\bottomrule
\end{tabular}
\label{TAB:symfun_values}
\end{center}
For training the HDNNPs of the LiOH system the RuNNer code~\cite{P4444,P5128} has been used. Various atomic NN architectures containing two or three hidden layers with 10 to 30 neurons per layer have been tested. The number of nodes was chosen to obtain a funnel-like architecture, i.e., each hidden layer contains less neurons than the previous hidden layer. The highest accuracy was found employing an architecture containing 25 nodes in the first hidden layer, 20 nodes in the second hidden layer, and 15 nodes in the third hidden layer resulting in a number of 3251 weights per element. The values of these weights were initialized following a modified Xavier initialization~\cite{P5866}.
\end{mdframed}

\section{Training}\label{sec:training}

\subsection{Optimization of the weight parameters \label{sec:optimization}}

The aim of the training process is to minimize the deviations between the HDNNP and the reference PES in the energetically relevant range. This is achieved by adjusting the values of the weight parameters using information from the reference data set. 
The optimization of the atomic NNs is a high-dimensional problem depending on thousands of weight and bias parameters and consequently it is impossible to find the global minimum in this parameter space. Still, there is a large number of roughly equivalent local minima of high quality, which represent the training data with low root mean squared errors (RMSE, see Box 4).
\\
\begin{mdframed}[style = mystyle,frametitle = {
Box 4: Error calculation
}]
\label{box:error}
The quality of the potential is continuously monitored by computing the root mean squared error (RMSE) of the energies of all structures
\begin{align}
E\; \mathrm{RMSE}=\sqrt{\frac{1}{N_{\mathrm{struct}}} \sum_{i=1}^{N_{\mathrm{struct}}}\left(E_{i, \mathrm{Ref}}-E_{i, \mathrm{HDNNP}}\right)^2}
\end{align}
and more rarely by the mean absolute deviation (MAD) or mean absolute error (MAE), which is typically smaller than the RMSE,
\begin{align}
E\; \mathrm{MAD}=\frac{1}{N_{\mathrm{struct}}} \sum_{i=1}^{N_{\mathrm{struct}}}\left|E_{i, \mathrm{Ref}}-E_{i, \mathrm{HDNNP}}\right| \quad .
\end{align}
For the force RMSE, there are two possible definitions, which are both in use. The first refers to the force components of all atoms in all structures,
\begin{align}
  F\; \mathrm{RMSE}=\sqrt{\frac{1}{N_{\mathrm{comp}}} \sum_{i=1}^{N_{\mathrm{comp}}} \left(F_{i,\mathrm{Ref}}-F_{i,\mathrm{HDNNP}}\right)^2} ,
\end{align}
while the second is computed as the error in force vectors between the HDNNP and the reference method,
\begin{align}
  F\; \mathrm{RMSE}=\sqrt{\frac{1}{N_{\mathrm{atoms}}} \sum_{i=1}^{N_{\mathrm{atoms}}}  \left(\mathbf{F}_{i,\mathrm{Ref}}-\mathbf{F}_{i,\mathrm{HDNNP}}\right)^2} ,
\end{align}
for all atoms in the data set. The force RMSE with respect to the vectors is larger by a factor $\sqrt{3}$ compared to the component-based RMSE. 
\end{mdframed}

\begin{figure}[!ht]
    \centering
    \includegraphics[width=\columnwidth]{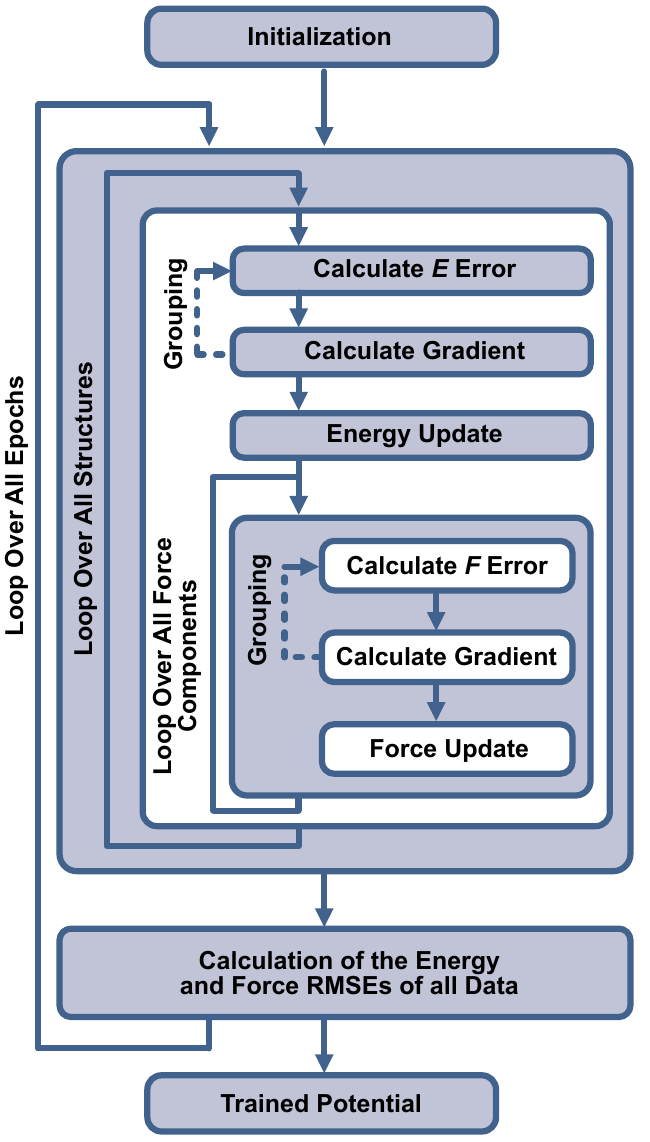}
    \caption{Flowchart of the weight optimization process using energies and forces. In a point-by-point update strategy the weights are updated once per energy and per force component, while also batch-wise grouping of errors and gradients is possible. At the end of each epoch, the RMSEs of the entire training and also test sets are computed.
    \label{fig:pointbypoint}}
\end{figure}

\begin{figure*}[!ht]
    \centering
    \includegraphics[width=2\columnwidth]{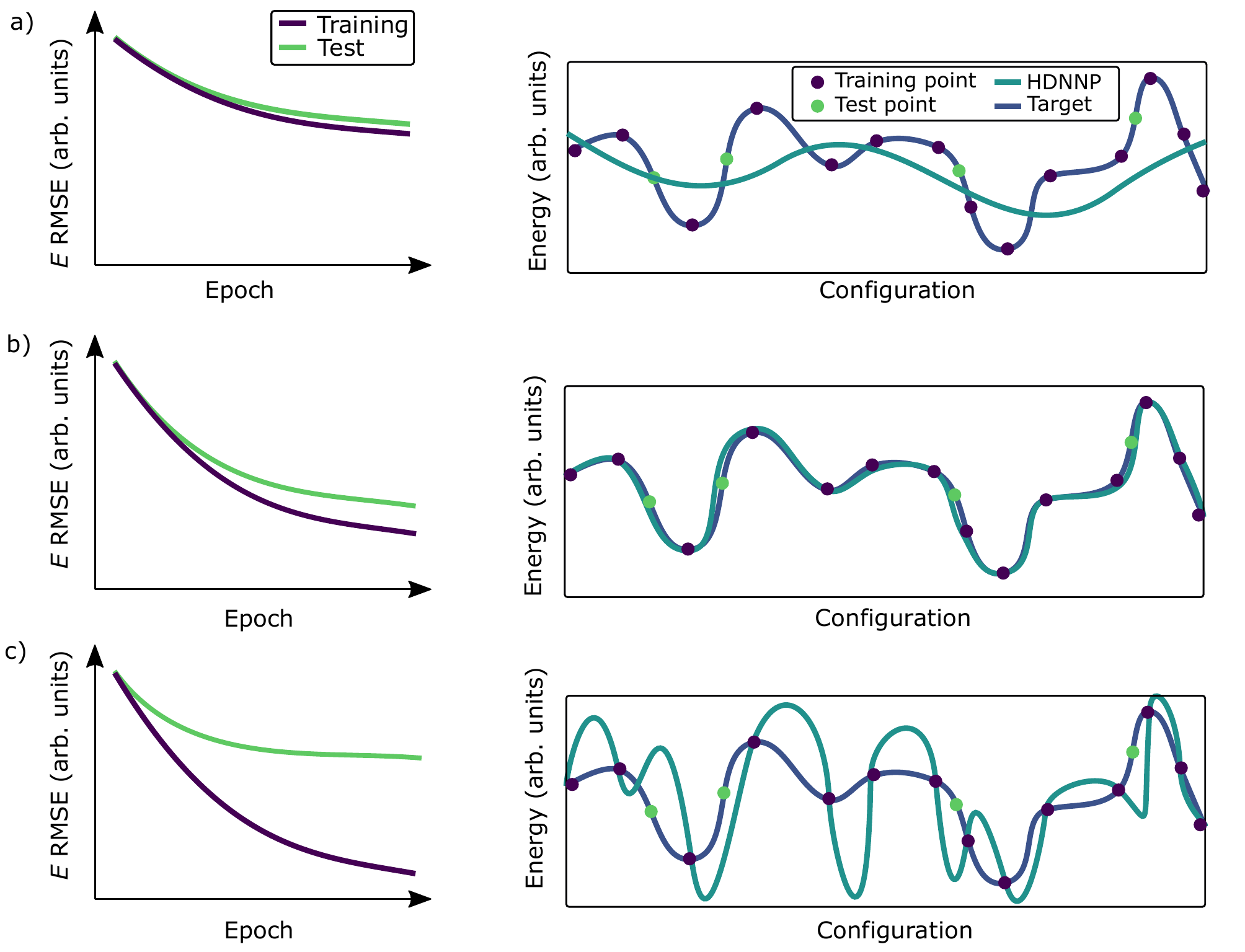}
    \caption{Illustration of overfitting. The predictive power for structures of the target surface not included in the training can be estimated by comparing the root mean squared errors (RMSE) of the training set and an independent test set. Panel a) shows a typical HDNNP for the case of ``underfitting'' due to the use of small and thus inflexible neural networks. Both, the training and the test set RMSEs converge to high values (left) and neither the training nor the test points are well represented (right). The fine details of the potential energy surface cannot be reproduced and only rough features can be learned. Panel b) displays the RMSEs obtained when using more flexible neural networks, which can represent all details present in the training set. Also the test points are reliably predicted if they are not too different from the training data and if the flexibility is kept as low as necessary. The case of too flexible neural networks is shown in panel c). Here, the RMSE of the training set is very low, and as can be seen on the right, all training points are closely matched although the rapid oscillations resulting from the high flexibility  prevent accurate predictions for the test data. This ``overfitting'', which can be detected by comparing the RMSEs of the training and the test set (``early-stopping method''), can be reduced by also including gradient information, i.e., the forces, of the training points.}
    \label{fig:overfitting}
\end{figure*}

The typical order of magnitude of the energy and force RMSEs of high-dimensional MLPs is about 1~meV/atom and 0.1~eV/\AA{}, respectively, but these values strongly depend on the diversity of atomic configurations, the quality of the data set and the energy and force ranges to be represented. The energy error is usually normalized per atom to make the RMSE size-consistent. For instance, a primitive unit cell of an fcc metal containing only one atom should have the same energy error as the conventional cubic unit cell containing four atoms, since the structures are identical.
\\
The training is done iteratively presenting energies and forces of the training data again and again for a predefined number of epochs (iterations), i.e., by supervised learning. However, the training process contains much more than merely the optimization of the parameter values. Instead, it also includes a continuous assessment of the quality of the potential (see Sec.~\ref{sec:validation}), and the extension of the data set by active learning (see Sec.~\ref{sec:activelearning}).
\\
Typical data sets of about 10,000 structures, which may contain roughly 100 atoms each, provide a wealth of information, i.e., 10,000 total energies and about 1,000,000 force vectors for each atom in the system, amounting in total to 3,010,000 pieces of information that can be exploited for the weight optimization. This is orders of magnitude larger than the number of parameters in HDNNPs, but this relation has to be treated with care as the formal amount of data does not contain any information about the diversity of the data set and its suitability to cover the configuration space of interest. For instance, the energy and force RMSEs will be low although the overall shape of the PES is not reliably represented if only a small subspace is sampled and all structures are very similar. Therefore, validating the PES quality for diverse structures, which are representative for the intended applications, is essential, and RMSE values alone can be difficult to interpret, because RMSEs can be arbitrarily low when computed for homogeneous and strongly correlated data sets.
\\
The course of the training process is shown schematically in Fig.~\ref{fig:pointbypoint}. Starting with an initial set of weights (see Sec.~\ref{sec:weightsanddata}), in each epoch the optimization algorithm loops over all structures in the training set in random order. There are different ways to update the weights, batch-wise for a group of energies or forces, which may even comprise the entire data set, or after the presentation of each individual piece of information in a point-by-point fashion corresponding to a batch-size of one. 
\\
Since in a point-by-point training strategy the parameters are updated once for each energy and for each force component, an epoch contains a large number of weight adjustments resulting in a rapid progress of the training in terms of epochs, while numerous updates also increase the computational costs of each of these epochs. In particular in case of large numbers of force components such a procedure can be very time-consuming. 
\\
To speed-up the training process, several measures can be taken. First, in adaptive training algorithms~\cite{P1308} an error threshold is defined for the energies and forces, and only the energies and forces exceeding this threshold are used for the training, while those, which are already accurately represented, are skipped. A common procedure is to define the energy and force error thresholds in terms of the respective RMSEs of the previous epoch such that the update criteria become tighter along with the improvement of the potential. As a consequence, in initial epochs when the RMSEs are still high only a few data points will be used and the overall shape of the PES will be learned quickly, while in later stages more information will be processed resulting in increasing computational costs of later epochs when the fine details of the PES are learned.
\\
Another optional way of reducing the number of weight updates is to use batches of energy and forces and to perform an update only every $N^{\mathrm{th}}$ energy or force component or even structure. In this approach, the individual errors and their gradients are accumulated, finally averaged, and used only for a single weight update per data group, which is particularly suited for parallelization. Such a parallelization is more difficult in case of constantly changing weight parameter values in the point-by-point approach in analogy to MD simulations, in which several MD steps cannot be parallelized due to the changing positions. Still, the latter update strategy is often more efficient, and reasonable potentials can be obtained after about 30-50 epochs, while in case of large batches that might even contain the entire training set thousands of epochs may be needed to reach a satisfactory accuracy of the potential.
\\
The most basic loss function $\Gamma_{\mathrm{E}}$ to be minimized in the training for a set of $N_{\mathrm{batch}}$ energies is given by 
\begin{align}
\Gamma_\text{E}=\frac{1}{N_{\mathrm{batch }}} \sum_{i=1}^{N_{\mathrm{batch }}}\left(E_{i,\mathrm{Ref}}-E_{i,\mathrm{HDNNP}}\right)^{2} \;.   
\end{align}
It should be noted that although the output values of the individual atomic NNs can be interpreted as atomic energies, these are just mathematical auxiliary quantities without a physical meaning. Hence, $\Gamma_{\mathrm{E}}$ is constructed using total binding energies only, without prior partitioning into individual atomic energies.
For gradient-based optimization algorithms the value and the derivative of this loss function with respect to the HDNNP parameters have to be computed.  The analytic derivatives are efficiently available by recursion through backpropagation~\cite{P2414}. Libraries for the construction of neural networks such as TensorFlow~\cite{P6541} or PyTorch~\cite{P6546} directly provide access to these gradients.
\\
The loss function for the forces is defined in a similar way, but here the batch of $N_{\mathrm{batch}}$ force components $F_i$ over all atoms of all structures is evaluated,
\begin{align}
    \Gamma_\text{F}=\frac{\beta}{N_{\mathrm{batch}}} \sum_{i=1}^{N_{\mathrm{batch}}}\left(F_{i,\mathrm{Ref}}-F_{i,\mathrm{HDNNP}}\right)^{2}  \quad.
\end{align}
\\
The constant prefactor $\beta$ can be used to balance the relative impact of the energies and forces in a joint loss function $\Gamma_{\mathrm{E}}+\Gamma_{\mathrm{F}}$, but for instance in point-by-point update strategies such a joint loss function is not required.
\\
Once the loop over all structures as well as their energies and forces has been completed and the respective updates of the weight parameters have been carried out, finally the new RMSE values of the energies and forces in the training as well as in the test set are computed and stored to assess the improvement of the potential. Then, the next epoch is started until the selected number of epochs has been completed and the final potential is obtained, which needs to be further validated (see Sec.~\ref{sec:validation}).
\\
The weight updates are usually performed using gradient-based optimization algorithms, such as the global adaptive extended Kalman filter~\cite{P0425,P1308,P4610} (see Box 5), conjugate gradients~\cite{Hestenes1952MethodsOC}, or the  Levenberg-Marquardt algorithm~\cite{P5263}. Very simple algorithms, such as steepest descent are usually not used.
\\
Alternatively, for large data sets the computational performance can be improved using the method of stochastic gradient descent\;\cite{sraOptimizationMachineLearning2012}. To optimize the convergence of this procedure momentum can be added or the learning rate can be modified, leading to algorithms such as the Adaptive Gradient Algorithm\;\cite{duchiAdaptiveSubgradientMethods2011}, AdaDelta\;\cite{zeilerADADELTAAdaptiveLearning2012} and the popular ADAM\;\cite{P5404} optimizer. 
\\
An important decision for the accuracy of the potential that has been made in the training process is the architecture of the atomic NNs, which governs the flexibility of the HDNNP. While too small networks prevent an accurate representation of the PES, too many parameters can give rise to overfitting, which can substantially reduce the transferability of the potential for structures not included in the training set. 
The presence of underfitting or overfitting can be monitored by the early-stopping method. For this purpose, the available reference data set is split into a training set used for the optimization of the weight and bias parameters, and an independent test set including only structures unknown to the HDNNP. Then, the error of both data sets in the iterative training process is observed (see Fig.~\ref{fig:overfitting}). If both, training and test set, do not reach low errors (Fig.~\ref{fig:overfitting}a), the flexibility of the HDNNP is not large enough. If the flexibility of the HDNNP is increased, both errors decrease resulting in a reasonable fit quality for both, known and unknown structures (Fig.~\ref{fig:overfitting}b). For too large atomic neural networks, overfitting results in a close to perfect fit for the known training data, while artificial oscillations are introduced, which yield poor predictions for the test data (Fig.~\ref{fig:overfitting}c). The reliability of the early-stopping method can be further improved by using cross-validation techniques and different splittings of the reference data, such that individual outliers are less relevant. Still, while useful for the determination of the required atomic NN architecture, the early-stopping method has to be used with great care as a tool to assess the overall quality of a potential as will be discussed in Sec.~\ref{sec:validation}.
\\
Even with an optimal architecture overfitting cannot be completely avoided. Still, it can be further reduced by the use of regularization methods such as ridge regression. As large values of the weights can be connected to overfitting, in ridge regression a term that penalizes large weight parameter values is added to the loss function. Another possibility to avoid overfitting is to use dropout \cite{P5691}. In this method nodes of the neural network are randomly dropped during training such that different thinned networks are trained. In this way complex co-adaptions are avoided in which layers would learn to correct mistakes of previous layers leading to bad generalization properties of the network.

\begin{figure}[!ht]
    \centering
    \includegraphics[width=\columnwidth]{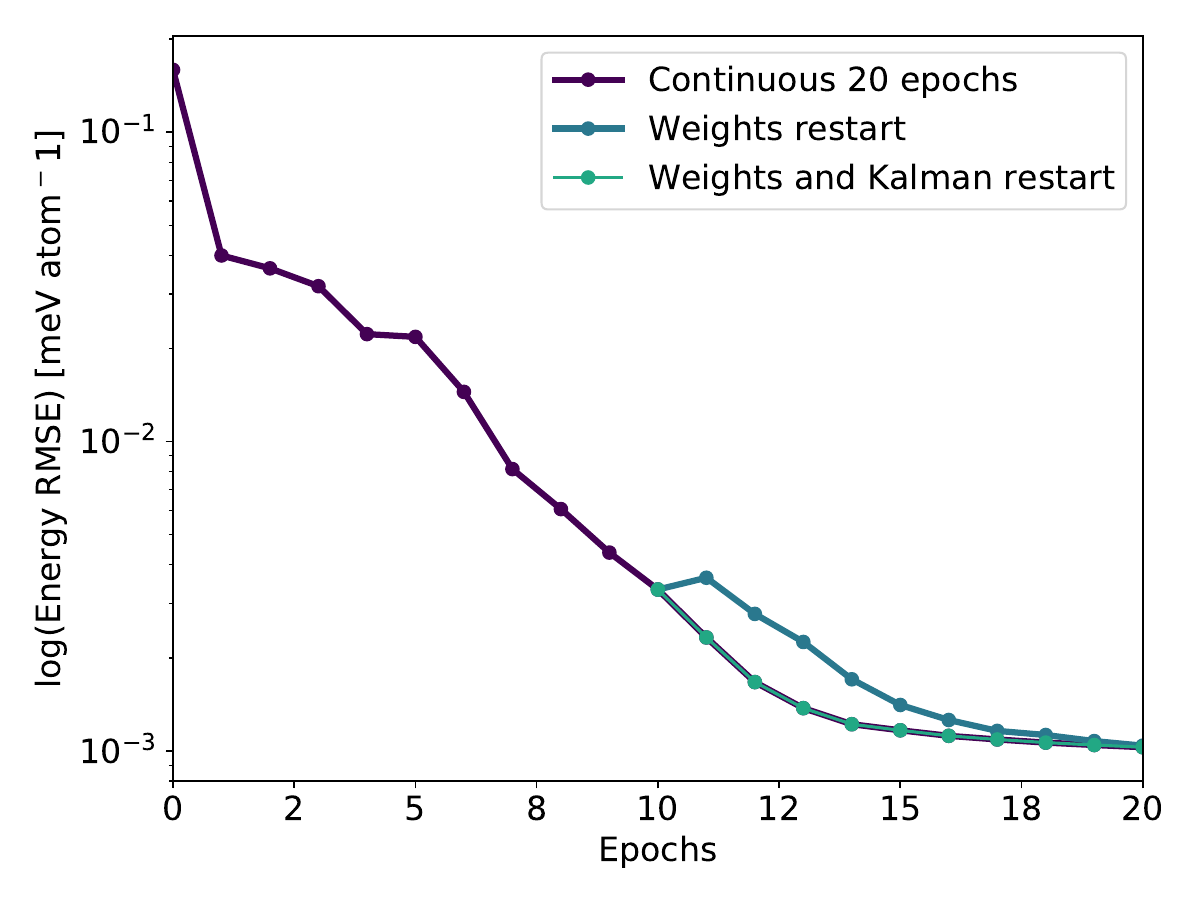}
    \caption{Restarting the training process with the Kalman filter~\cite{P1308}. The violet curve shows a typical decrease of the energy RMSE in a training process consisting of 20 epochs. If however, the training is stopped after 10 epochs and shall be restarted, there are two options. If only the weight parameters of epoch 10 are used in the restart, the further progress of the training differs from the violet reference curve, since the covariance matrix of the Kalman filter is initialized again from scratch. If, however, this Kalman matrix is stored and used in the restart along with the weights, the obtained RMSE values are indistinguishable from the continuous fit of 20 epochs. 
   }
    \label{fig:restart}
\end{figure}

Finally, several points have to be considered when restarting the training process from parameters obtained in a previous optimization. Since these parameters have been determined for a well-defined set of structures and their associated ACSF values, any scaling factors possibly employed in the preprocessing of the ACSFs become part of the potential and need to be applied in the same way to newly added structures. If, for instance, the training set is increased before restarting the training process, these scaling factors change. Consequently, the scaled symmetry function values are modified resulting in slightly different energies and forces even for the structures included in the original data set and even if no adjustments in the weights have been made. This can be avoided by restarting the training process using fixed scaling factors at the cost of slight changes in the effective range of the ACSF values. Moreover, some optimization algorithms, such as the Kalman filter generate further information such as a covariance matrix in the fitting procedure, which must be stored are reused when restarting fits if a strict continuation of the training is intended (see Fig.~\ref{fig:restart}).

\begin{mdframed}[style = mystyle,frametitle = {
Box 5: Kalman Filter
}]
\label{box:kalman}
Originally, the Kalman filter has been developed to recursively find the minimum variance estimates of state variables of linear dynamic systems, which has been generalized in the extended Kalman filter to non-linear systems\;\cite{P0425}. Feed-forward neural networks can be considered as such non-linear dynamic systems and the Kalman filter can be efficiently used to find optimal weight parameters while recursively looping through a set of reference data\;\cite{P1308}.
\\
Generally, the update procedure is the same for all properties, which will be denoted here as $V$. According to the scheme in Fig.\;\ref{fig:pointbypoint}, the Kalman filter is usually used for a batch size of one, i.e., a weight update is performed after each presented energy or force component, respectively.
\\ 
Each update $k$ follows same procedure: First, the error is estimated as the absolute difference between the predicted target value $V_k$ and its reference value $V_k^{\mathrm{Ref}}$ and normalized by the number of atoms in the current structure $N_{\text {atoms}}$ 
\begin{align}
\nu_{k} = |V_k - V_k^{\mathrm{ref}}|N_{\text {atoms}}^{-1}\;.
\end{align} 
In the following adaptive step this error is compared to an error threshold, which must be exceeded to continue with the update process.\\ 
Then, the Kalman gain matrix $\mathbf{K}$ is computed using the Kalman covariance matrix of the previous step $\mathbf{P}_{k-1}$, the current value of a forgetting factor $\lambda$, the identity matrix $\mathbf{I}$, and the Jabobian $\mathbf{J}$ containing the derivatives of the loss function with respect to all weight parameters, 
\begin{align}
\mathbf{K}_{k}=\mathbf{P}_{k-1} \mathbf{J}_{k}\left[\lambda_{k-1} \mathbf{I}+\left(\mathbf{J}_{k}\right)^{T} \mathbf{P}_{k-1} \mathbf{J}_{k}\right]^{-1}\;.
\end{align}
The covariance matrix is usually initialized as the unity matrix. During the training, it stores information about all updates. Its diagonal elements can be interpreted as the uncertainty of the current weight estimate.
The gain matrix is then used to update the weight matrix $\mathbf{w}$ as
\begin{align}
\mathbf{w}_{k}=\mathbf{w}_{k-1}+\mathbf{K}_{k}  \nu_{k}\;.
\end{align}
In the following the covariance matrix and $\lambda$ are updated
\begin{align}
\mathbf{P}_{k}=\lambda_{k}^{-1}\left[\mathbf{I}-\mathbf{K}_{k}\left(\mathbf{J}_{k}\right)^{T}\right] \mathbf{P}_{k-1}\;.
\end{align}
\begin{align}
\lambda_{k}=\lambda_{k-1} \lambda_{0}+1-\lambda_{0}\;.
\end{align}
$\lambda$ here takes the role of keeping the information of previous updates in the current update. Its initial value $\lambda_0$ is chosen close to unity and its value increases as the optimization progresses. As the value of $\lambda$ increases, the amount of reference date points considered in each update also increases. This has the advantage to enable larger changes at the beginning of the optimizing process and thereby avoiding to get stuck in local minima. When the optimizing proceeds, more information is used and the local minima can be found by taking smaller steps.
\end{mdframed}

\subsection{Active Learning \label{sec:activelearning}}

\begin{figure*}[t!]
    \centering
    \includegraphics[width=2\columnwidth]{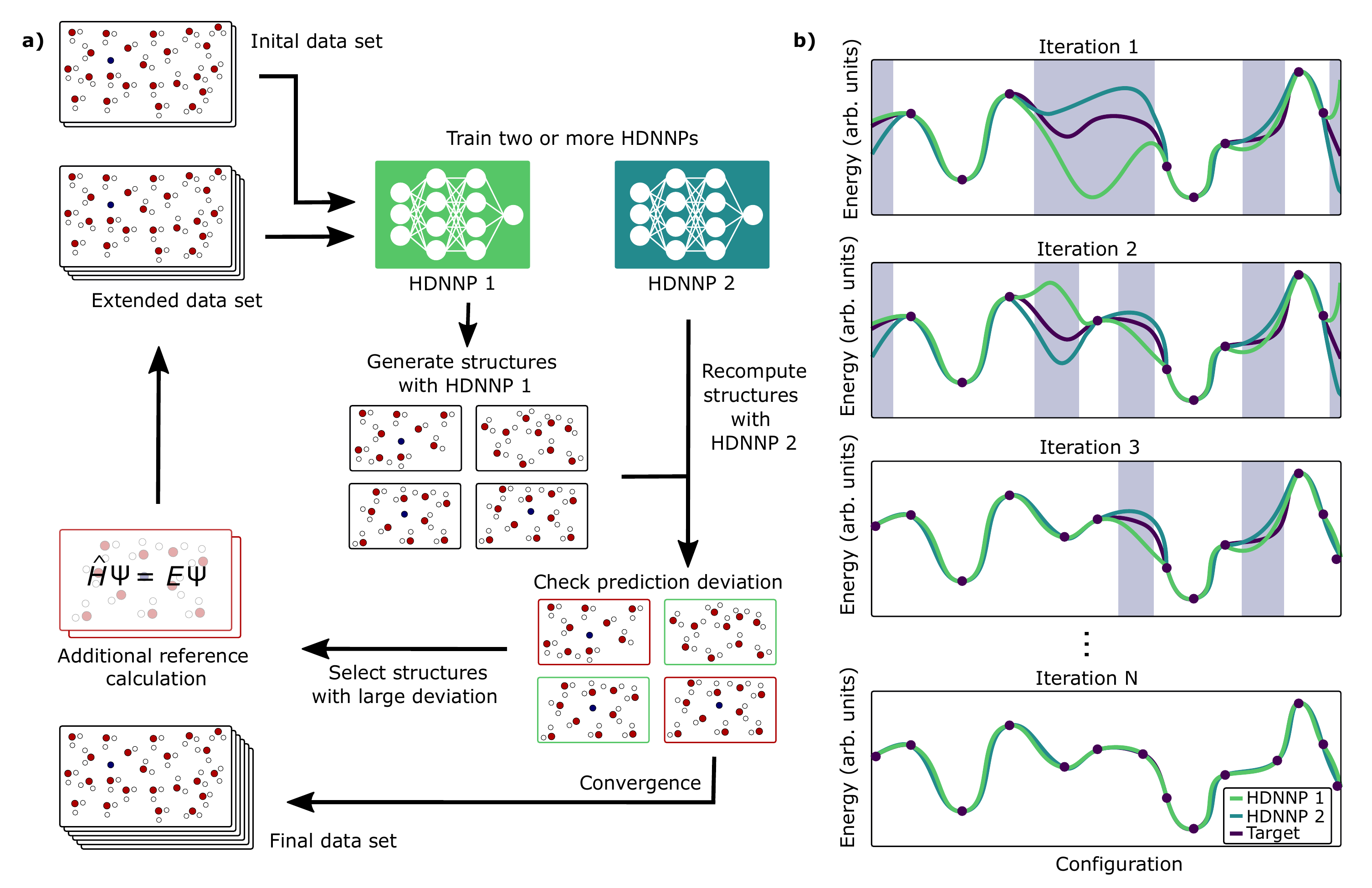}
    \caption{Iterative improvement of HDNNPs by active learning. Panel a) shows a flowchart of the self-consistent determination of the required training data set. Starting from an initial data set, several HDNNPs are trained. One of these HDNNPs is then used to generate a large pool of structures. These structures are then recomputed by the remaining HDNNPs. If the variance of predicted properties, such as energies or forces in the ensemble are too large for some structures (framed in red) these are selected for additional reference calculations to increase the data set. This extended data set is then used for training in the next cycle and so forth until no poorly predicted structures are found and a converged potential has been obtained. In panel b) this procedure is shown schematically for an unknown target potential energy surface (violet line). The symbols represent the available training points. Close to these training points HDNNP 1 and HDNNP 2 predict very similar energies, while the regions with high uncertainty (grey) decrease with increasing data set until the final HDNNPs represent the potential energy surface in the entire region of interest.
    }
    \label{fig:activelearning-PES}
\end{figure*}

\begin{figure}[t!]
    \centering
    \includegraphics[width=\columnwidth]{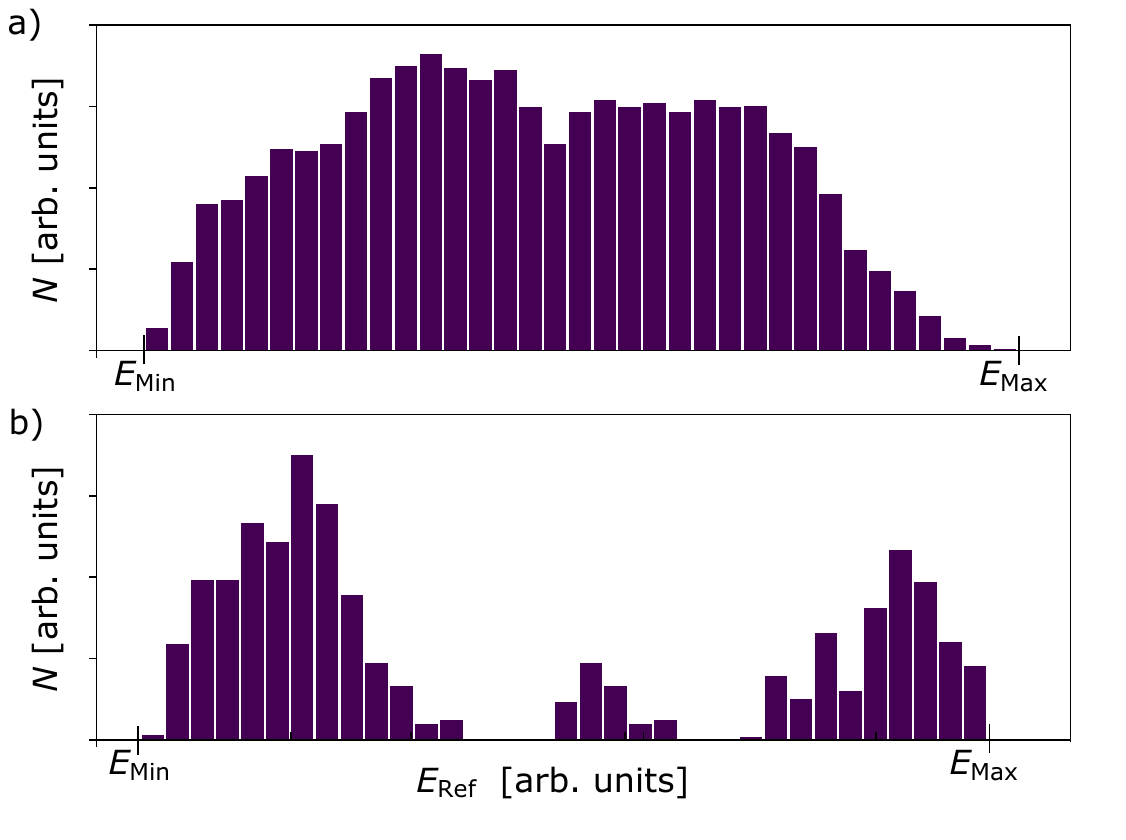}
    \caption{Possible shapes of energy histograms of the reference data (schematic). Panel a) shows a histogram with balanced energy distribution in the energy range of interest. The lowest energy $E_{\mathrm{min}}$ is the energy of the optimized system in its global minimum configuration. $E_\mathrm{max}$ is the highest energy in the data set, which should be larger than the highest energy of interest. Panel b) shows an incomplete distribution of reference structures. Such isolated groups of structures can point to disconnected parts of trajectories in configuration space. These disconnected regions must be filled in the active learning process if transitions between these structure types are relevant for the intended simulations. The energies are normalized per atom to avoid offsets due to possible different chemical compositions.}
    \label{fig:Ehistogram}
\end{figure}

Since the quality of the potential obtained in the training process critically depends on the available data set, an important task in the construction of MLPs is the generation of reference data sets covering the relevant configuration space as comprehensively as possible. While in conventional empirical potentials and classical force fields a certain transferability is ensured by the physically inspired functional form, the extrapolation capabilities of MLPs are very limited and all information about the topology of the PES must be learned from the reference data. Therefore, the reference structures must cover the part of the PES, which is energetically accessible in the intended simulations. For the data set, not only conditions such as temperature and pressure, but depending on the intended application also different structures such as various polymorphs of crystal in case of solids or configurations occurring in classical and quantum simulations of liquids such as water~\cite{P4971,P6360} need to be considered. Further, repulsive structures for close atomic encounters need to be learned, which can also be assisted by explicitly including two- and three-body terms~\cite{P6021,P6542}. 
\\
Since there is no hope to have at hand a complete set of all relevant structures before the training process, an automatic and unbiased approach to identify important atomic configurations is needed. Ideally, these structures should be determined taking the current training status of the MLP and the existing data into account.
This approach is called active learning and is based on the concept of query by committee~\cite{P5900}. In this method, the high flexibility of MLPs, which is essential for their numerical accuracy but also the origin of failures for predictions far from the training geometries, is turned into an advantage by making use of the energy or force variance of predictions in an ensemble of MLPs when applied to unseen structures. To ensure that these structures are relevant for the intended applications of the MLP, they are often generated using similar simulation protocols and conditions including temperature and pressure as the final production simulations. If this variance is in the order of the RMSEs, then the prediction can be considered as reliable and the unseen structure is sufficiently close to a known training point. If, however, the variance is much larger, the unseen structure is far away from the training data and should be added to improve the potential in this region. In this way, step-by-step the data set and the quality of the MLP can be iteratively improved until a reliable potential has been generated. As a side note we mention that the use of a committee is not always necessary in all types of MLPs, as, for instance, in Gaussian process regression a single instance of the regressor can provide an estimate of the variance.
\\
The general procedure of active learning, which has become a standard in the development of modern MLPs~\cite{P6543,P3114,P4939,P5399,P5782,P5814,P5842} is shown in Fig.~\ref{fig:activelearning-PES}. First, an initial data set is generated (see Sec.~\ref{sec:initialdataset}) and an ensemble of MLPs is trained as represented by two HDNNPs in Fig.~\ref{fig:activelearning-PES}a. All members of the ensemble should have approximately the same quality in representing the available training data, as measured, e.g., by the energy and force RMSEs. Then, one of these HDNNPs is chosen to generate a large number of trial structures, e.g. by the simulation method of interest, and the second HDNNP is used to recompute the energies and forces for these structures. Alternatively, an ensemble of HDNNPs can be used to predict these properties on-the-fly during the simulation. Then, the variance in the predictions is investigated and structures with large deviations are selected for additional reference electronic structure calculations. These new data points are then added to the extended data set and used to train the next generation of MLPs. This procedure is repeated for several cycles until the variance of all trial structures remains close to the RMSEs. The HDNNPs trained to this final data set are then ready for the production simulations. As can be seen in Fig.~\ref{fig:activelearning-PES}b, adding more and more data points reduces the gray-shaded regions of high uncertainty in the PES until a reliable representation of the PES has been obtained.
\\
Several comments should be made at this point. While in particular for Gaussian approximation potentials the retraining of MLPs on-the-fly during MD simulations has been suggested~\cite{P5825}, this procedure is usually not applied in case of HDNNPs. Instead, for neural network potentials often a few hundred new structures are first identified before a new generation of potentials is trained. Before this training step, new structures should be carefully investigated since in particular in early stages of active learning unphysical structures may be suggested by poor preliminary potentials. Such structures complicate the training process while they are irrelevant for the final potential. Moreover, including similar structure should be avoided, which can be achieved, e.g., by farthest point sampling~\cite{Bartok2017}. 
\\
In general, monitoring the variance of the atomic forces should be preferred for the selection of new structures, since total energies are difficult to interpret in particular for large systems. The reason is the tendency for error compensation among the individual atomic energies, which can lead to apparently reliable energy predictions but in fact reduce transferability. Moreover, the investigation of atomic forces allows to identify specific atomic environment, which are not well represented in the training set. Transfering the respective atoms from complex systems to smaller structures for efficient reference calculations can be challenging, as a substantial part of the environment has to be included. For some systems the use of molecular fragments is a viable path to reduce the system size while keeping the relevant parts of the atomic environments intact~\cite{P4585,P5537,P6065,P6527}. Active learning and farthest point sampling can also be used to select to minimum number of structures from an already existing data set needed to achieve a certain accuracy of the potential~\cite{P5814}.
\\
The convergence of the data set in the active learning process can also be investigated by plotting histograms of the energy distributions in the training set. Fig.~\ref{fig:Ehistogram}a shows a balanced distribution of training structures starting from the smallest energy $E_{\mathrm{min}}$, which is often the optimized global ground state structure of the system, and the highest energy of interest $E_{\mathrm{max}}$. The possible presence of energy gaps, however, should be carefully investigated (Fig.\ref{fig:Ehistogram}b). Such gaps can arise from different chemical compositions of structures in the data set, which is in general no problem, but they can also point to missing parts of molecular trajectories. If, for example, a reaction from $A$ to $B$ shall be studied, a continuous representation of all intermediate structures and the respective energies in the data set is mandatory.

\begin{figure}[!ht]
    \centering
    \includegraphics[scale=0.80]{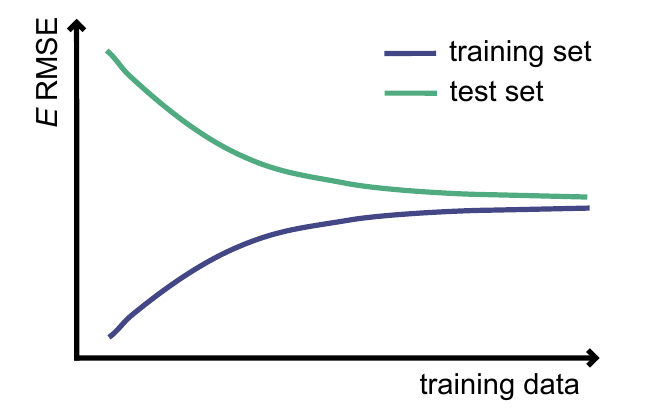}
    \caption{Schematic learning curve. Small training sets can be learned with high accuracy, but the resulting HDNNP has poor generalization capabilities and provides high errors for test structures not included in the training set. If the number of training structures is increased, the training process is more challenging resulting in an increased training error, while the test set error decreases due to the overall improving potential. For very large training sets, the configuration space is well represented by the training data and the root mean squared errors (RMSE) of the training and test sets are very similar. }
    \label{fig:learningcurve}
\end{figure}

Another common procedure to investigate the completeness of the data set is the use of learning curves (see Fig.~\ref{fig:learningcurve}). Here, the RMSEs of the training set and the test set are plotted as a function of the training set size. Small training sets can be accurately learned while they are not representative for the entire configuration space. Therefore, the RMSE of an independent test set is very large. If the training set is increased, the learning task is more difficult and the training error increases while the overall quality of the MLP improves such that the RMSE of the test set decreases. For an ideal, infinitely large data set, training and test set RMSEs converge to very similar values. However, the usefulness of learning curves strongly depends on the coverage of configuration space. If certain structures are completely missing, they are absent in both, the training and the test set, and consequently not true convergence can be achieved.

Several variants of active learning have been proposed to reduce the computational effort of the reference calculations. In $\Delta$-Learning~\cite{P2298,P3872,P6225,P4513} first a baseline potential is constructed. This baseline potential should be cost-effective to evaluate and can be, e.g., a simple empirical potential, a moderately expensive electronic structure method, or another MLP trained to a large data set obtained from such a method. This baseline potential is then used to represent the rough overall topology of the PES. In a second step a MLP potential is trained to a represent the energy difference between the baseline potential and a very accurate high-level electronic structure method. Since only a small energy range needs to be learned for this correction, its error is easier to control. Moreover, the hope is that a smaller data set may be required compared to a conventional MLP completely based directly on high-level data. Still, as usually the corrugation of the PES is more or less independent of the choice of the reference method, the possibility for the reduction of the data set size is limited. An interesting use of $\Delta$-Learning is the combination of a cost-effective method providing energies and analytic forces, and an expensive method for which no forces are available~\cite{P6360}. In this case, at least the baseline potential benefits from the availability of additional force information.
\\
Finally, another approach is transfer learning~\cite{P5621}. Here, first a MLP is trained to a large and computationally affordable reference data set. A part of the training data is then recalculated at a higher level of theory and the potential is retrained to this data set. Some parameter of the MLP are fixed and do not change during the retraining while others can adjust to represent the modified shape of the PES at reduced costs.

\begin{mdframed}[style = mystyle,frametitle = {Case study: Training of the LiOH-water system}]
The initial reference data set of the LiOH system was iteratively expanded following the active learning scheme. Six cycles of active learning were conducted using the tool RuNNerActiveLearn~\cite{P6057}. In Fig.\;\ref{fig:LiOH_data_hist} histograms of the reference data set of each cycle are shown.
\begin{center}
    \includegraphics[width=8cm]{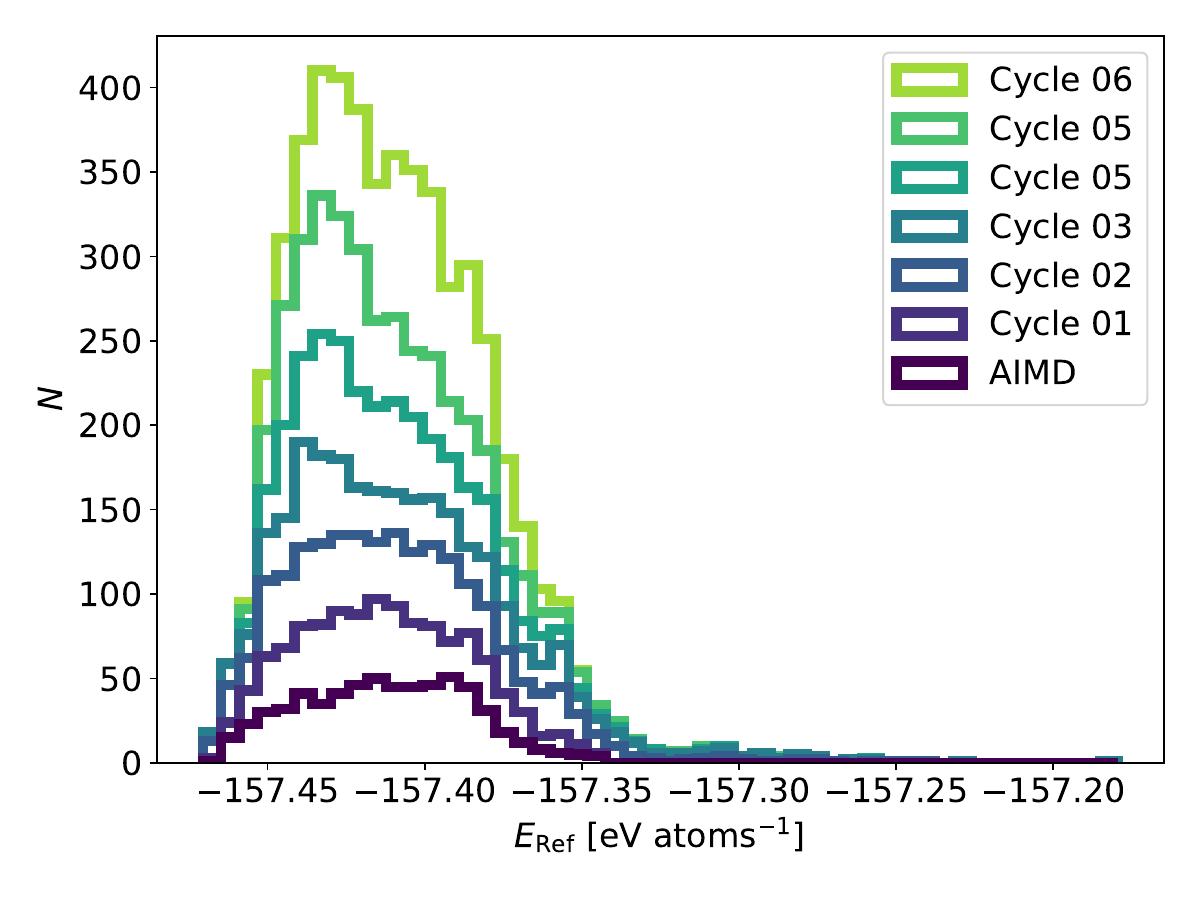}
      \captionof{figure}{Histograms of the lithium hydroxide data set at each cycle of the active learning process. The full data set also contains pure water structures, which are not included in these histograms for clarity. The initial data set has been generated by ab initio MD (AIMD).}
     \label{fig:LiOH_data_hist}
      \par
\end{center}

The progress of the active learning can also be monitored by evaluating the RMSE values after each completed cycle of adding new structures and retraining the potential. The resulting RMSE values are shown in Fig.~\ref{fig:lioh_learning}. The obtained curve represents a learning curve, but there are notable differences in the shape of test set RMSE values compared to the idealized learning curve shown in Fig.~\ref{fig:learningcurve}, which would be found when extracting data from a close-to complete data set that is never available in any realistic scenario for complex systems. In the real case of Fig.~\ref{fig:lioh_learning}, the data set increases from cycle to cycle and new parts of configuration space are explored, which thus increase the complexity in the training and the test set in the same way. Hence, the test points are described with comparable accuracy as the training points illustrating the need for a careful use of learning curves in case of incomplete data sets. If at all, learning curves are best used for an already large and converged data set starting from a small subset of data, which is then step-by-step increased, but they are not a very useful tool during active learning cycles.

\begin{center}
    \includegraphics[width=8cm]{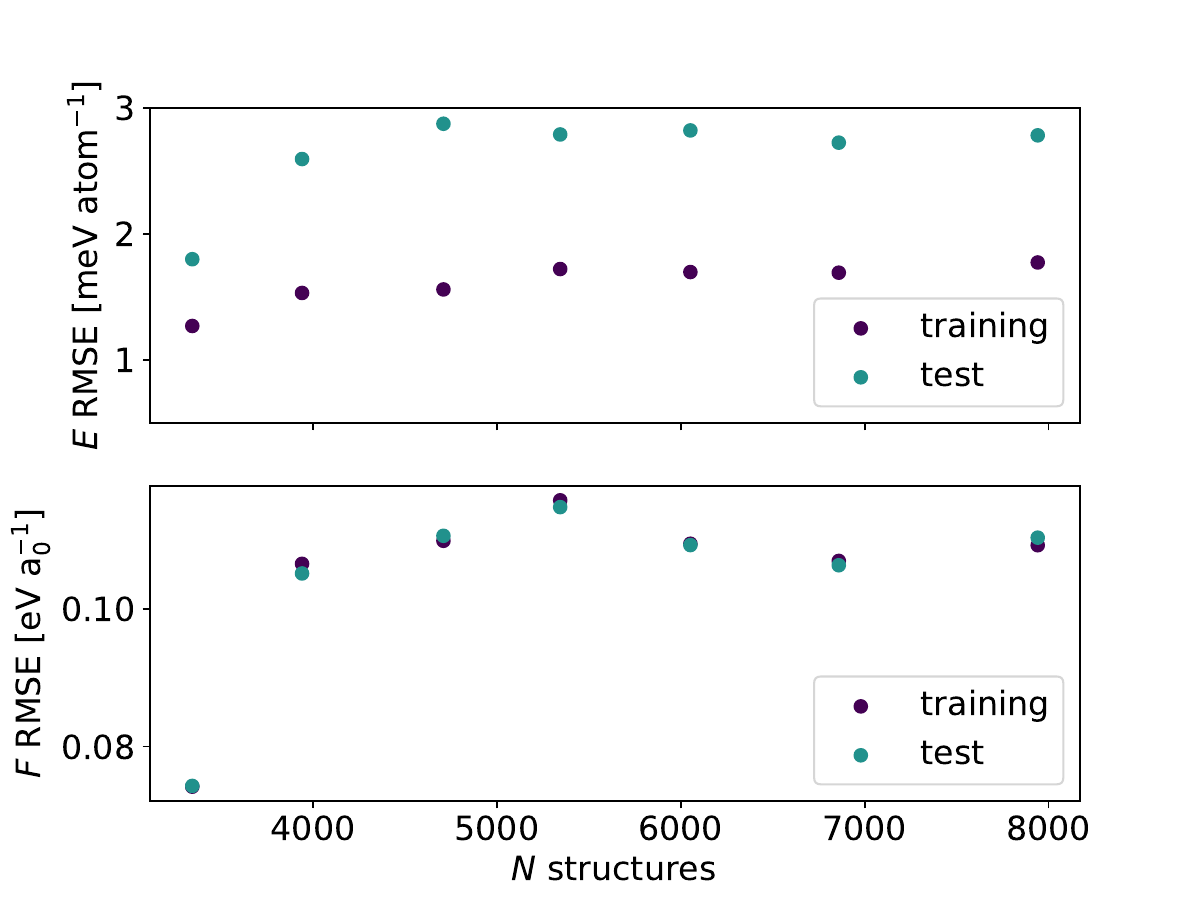}
      \captionof{figure}{Change of the RMSE values during the active learning cycles. For consistency the same settings and architectures of HDNNPs have been used in all potentials. The radial symmetry function $\eta$ values are adjusted to the current data set.}
     \label{fig:lioh_learning}
      \par
\end{center}

Overall, the most accurate HDNNP obtained for the final LiOH data set has energy RMSEs of $1.204\;\mathrm{meV\;atom^{-1}}$ for the training and $1.682\;\mathrm{meV\;atom^{-1}}$ for the test data set. The force component RMSEs are $0.069272\;\mathrm{eV\;a_0^{-1}}$ for the training and $0.069355\;\mathrm{eV\;a_0^{-1}}$ for the test data set.

\end{mdframed}

\subsection{Beyond Short-Ranged Potentials}\label{sec:beyondshortranged}

\begin{figure}[!ht]
    \centering
    \includegraphics[width=\columnwidth]{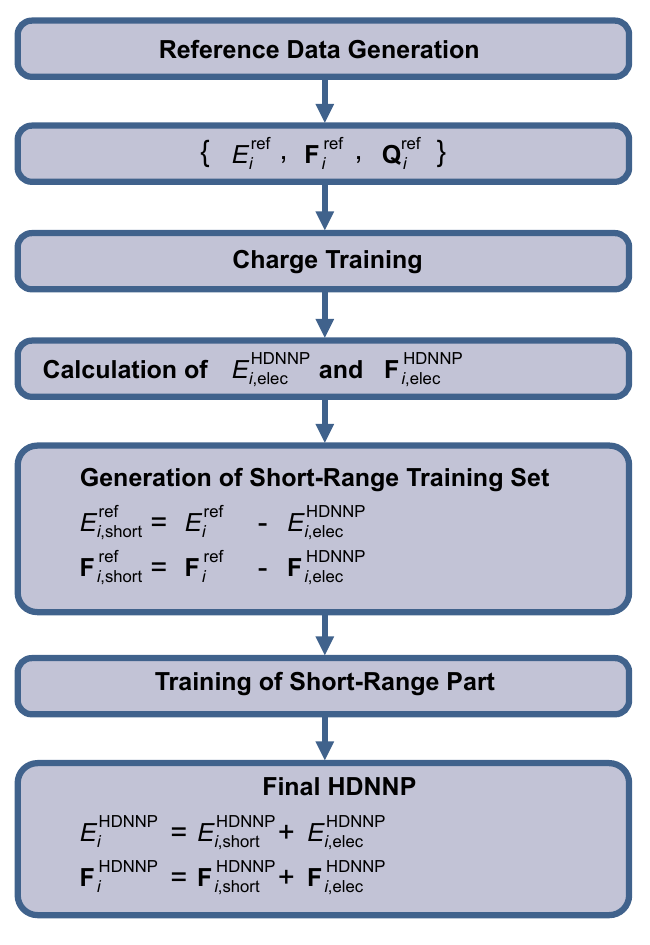}
    \caption{Training of HDNNPs including long-range electrostatic interactions. First, the reference energies, forces and atomic charges are obtained from electronic structure calculations. Then, the charges are trained and the HDNNP electrostatic energies and forces are computed. After applying a screening (Eq.~\ref{eq:screenfunction}) these are then removed from the reference energies and forces to obtain the target energies and forces for the training of the short-range part. Once also the short-range part has been learned, the total HDNNP is given as a sum of the short range and electrostatic energies and forces (Eq.~\ref{eq:eelec}).}
    \label{fig:flowchart-electrostatics}
\end{figure}

\begin{figure}[!ht]
    \centering
    \includegraphics[width=\columnwidth]{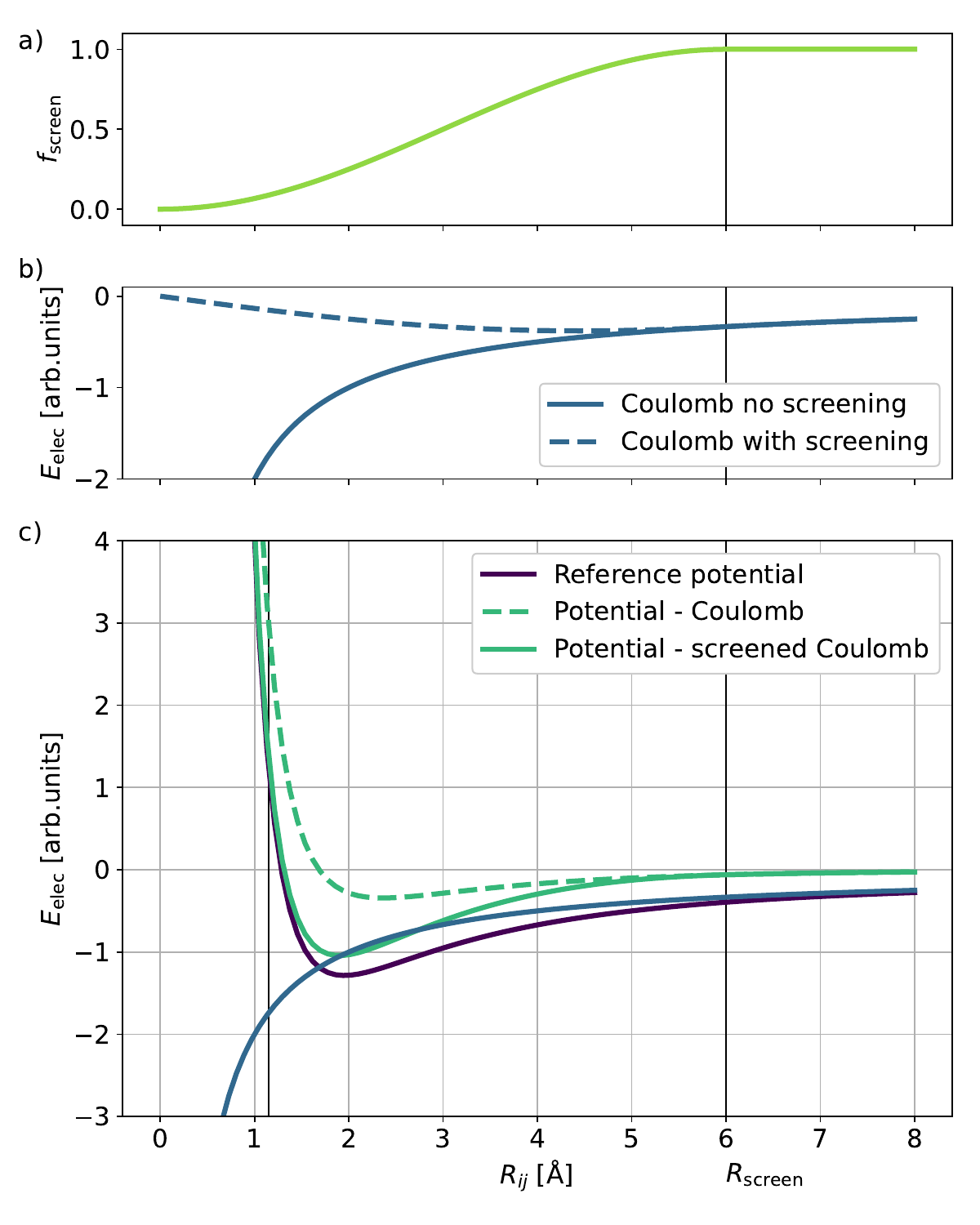}
    \caption{Screening of electrostatic interactions~\cite{P3132}. Panel a) shows the screening function $f_{\mathrm{screen}}$ (Eq.~\ref{eq:screenfunction}), which smoothly decays from one to zero inside the screening radius $R_{\mathrm{screen}}=6$\AA{}. Panel b) shows the Coulomb energy of two atoms with opposite charges as well as the screened Coulomb interaction, which approaches zero for short interatomic distances. Inside the cutoff radius of the ACSFs the missing part of the Coulomb energy can be represented by the short-range atomic energies.
    Panel c) shows a reference pair potential $E_{\mathrm{ref}}$, as well as the energy curves obtained when removing the unscreened Coulomb energy ($E_{\mathrm{ref}}-E_{\mathrm{Coulomb}}$) or the screened Coulomb energy ($E_{\mathrm{ref}}-E_{\mathrm{screened}}$). The curve obtained using the screened Coulomb potential covers a much smaller energy range and thus can be represented more accurately by the short-range energy. 
    }
    \label{fig:screening}
\end{figure}

So far, we have discussed all the steps of the construction of MLPs for the example of second-generation HDNNPs. Most of these steps are general and equally apply to third- and fourth-generation HDNNPs. However, the total energy expression of third- and fourth-generation HDNNPs consists of the electrostatic and the short-range parts (Eq.~\ref{eq:eelec}), which have to be trained sequentially following the flowchart in Fig.~\ref{fig:flowchart-electrostatics}.
\\
First, the energies, forces and atomic partial charges are determined in reference electronic structure calculations. Since atomic charges are not quantum mechanical observables, many different partitioning schemes can be used, such as Hirshfeld~\cite{P0416}, Bader~\cite{P1711} or density-derived electrostatic and chemical (DDEC) charges~\cite{P5395}. Any numerical uncertainty resulting from this choice can be compensated inside the atomic environments by the short-range atomic energies. Then, the charges are learned, which is done either by using a second set of atomic NNs (third-generation HDNNPs) or by training environment-dependent electronegativities (fourth-generation HDNNPs) yielding these charges in a charge equilibration process. Once an expression for the charges has been learned, the electrostatic energies and forces are computed and removed from the total reference energies and forces to determine the reference data for the short-range training. This sequential procedure has two reasons: first, a functional relation between the structure-dependent charges and the atomic positions needs to be available to compute the electrostatic forces, which contain the derivatives of the charges with respect to the atomic coordinates~\cite{P6018}. Such a relation cannot be obtained from the reference electronic structure calculations. Second, since all those energy contributions, which are not electrostatic, are combined in the short-range part, there is no double counting of electrostatics by construction. Finally, the remaining short-range energies and forces are trained to yield the HDNNP consisting of the electrostatic and the short-range part.
\\
The Coulomb potential has a singularity for short interatomic distances, which can give rise to an increased short-range energy range, which is more difficult to learn than the original reference energies. Therefore, the Coulomb interaction is usually screend to zero inside a screening radius $R_{\mathrm{screen}}$, which must be lower than the cutoff radius of the ACSFs as illustrated in Fig.~\ref{fig:screening}. The screening function~\cite{P3132} (Fig.~\ref{fig:screening}a) is given by
\begin{eqnarray}
\label{eq:screenfunction}
f_{\rm screen}=\begin{cases}
  \frac{1}{2}\Bigl[ 1-\cos\Bigl(\frac{\pi R_{ij}}{R_{\mathrm{screen}}}\Bigr)\Bigr]  & \text{for }R_{ij}\le R_{\mathrm{screen}}\\
  1 & \text{for }R_{ij}>R_{\mathrm{screen}} .
\end{cases}
\end{eqnarray}
When multiplied by this screening function, the Coulomb potential decays smoothly to zero for small distances $R_{ij}$ (Fig.~\ref{fig:screening}b). This screened Coulomb potential is then removed from the total energy reference curve before training the short-range part (Fig.~\ref{fig:screening}c). As can be seen for the example of an interatomic distance of about 1.2~\AA{}, removing the screened Coulomb potential results in a much smaller energy range to be learned by the short-range part compared to the case of removing the unmodified Coulomb potential.

\section{Validation}\label{sec:validation}

\begin{figure}[!ht]
    \centering
    \includegraphics[width=\columnwidth]{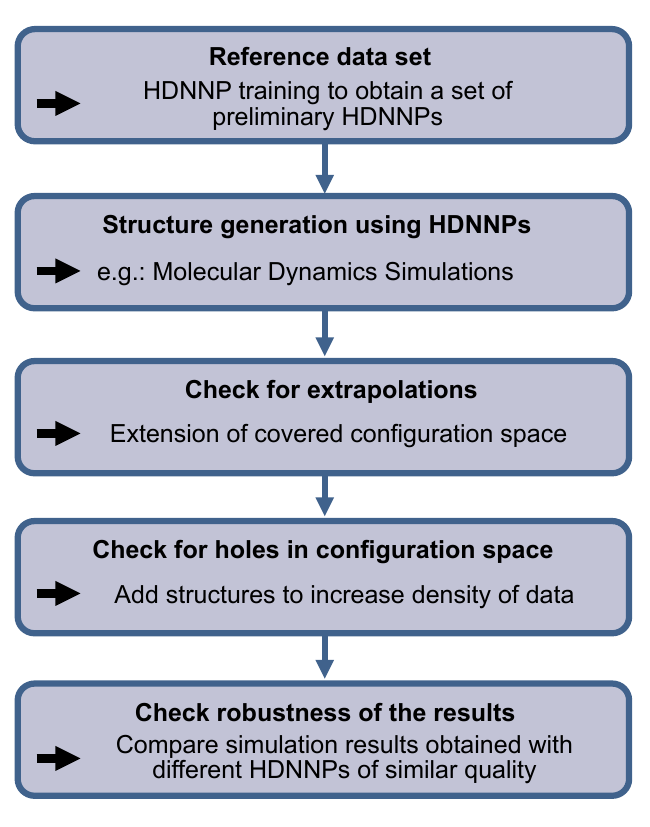}
    \caption{Multistep validation of machine learning potentials.}
    \label{fig:multistepvalidation}
\end{figure}

\begin{figure}[!ht]
    \centering
    \includegraphics[width=\columnwidth]{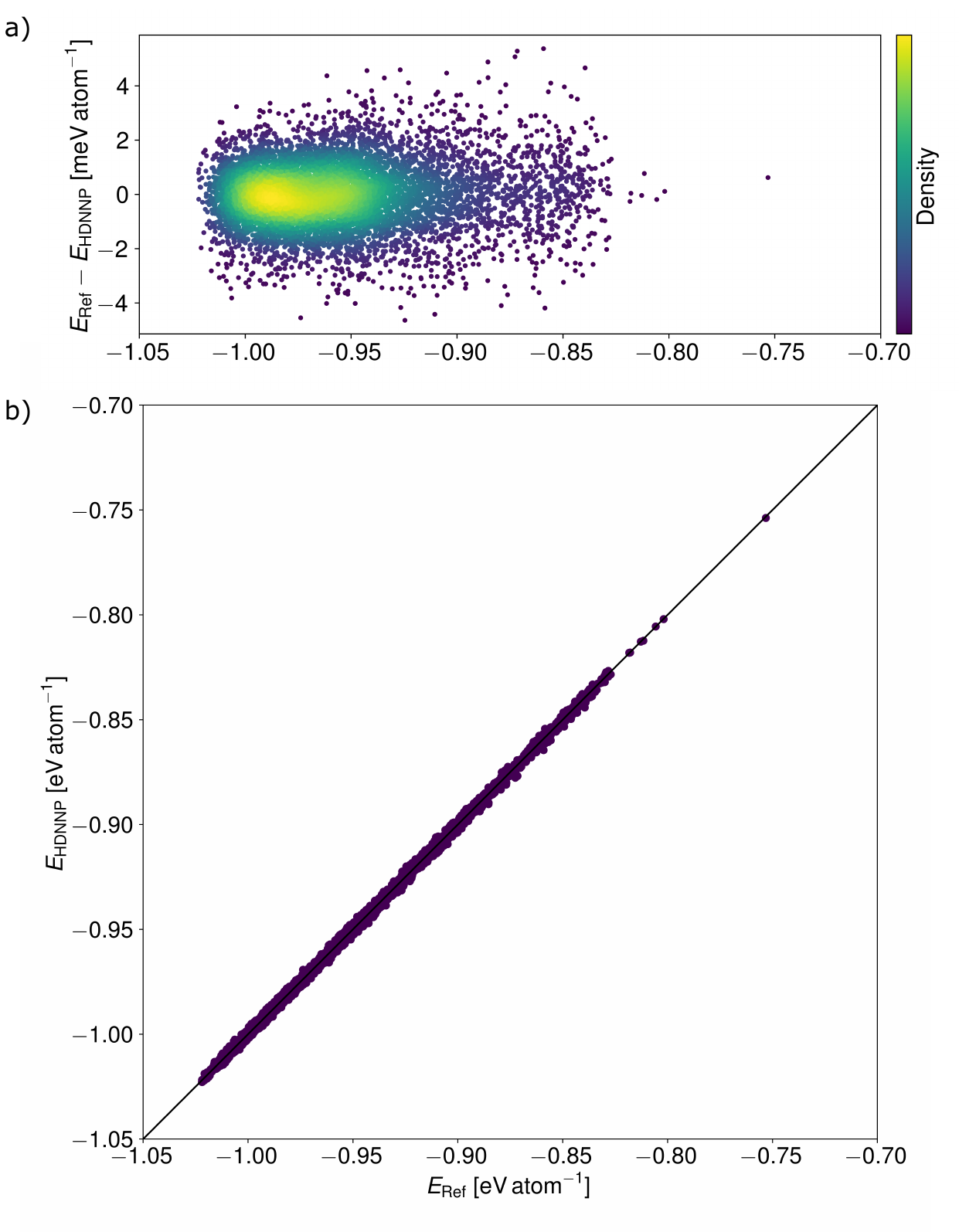}
    \caption{Energy correlation plots of the LiOH-water data set. Panel a) shows the signed energy errors and the density of data points as a function of the target energy of the reference method. Panel b) shows the correlation of the HDNNP energies and the energies of the reference method. To avoid energy offsets, binding energies are used for this purpose instead of total energies, which would strongly depend on the chemical composition. For a perfect potential, all points should be aligned along the diagonal line with slope of 45$^{\circ}$.
    Correlation plots should be generated separately for the training and the test data set for the energies as well as the force components. 
    }
    \label{fig:enn_eref}
\end{figure}

\begin{figure}[t!]
    \centering
    \includegraphics[width=\columnwidth]{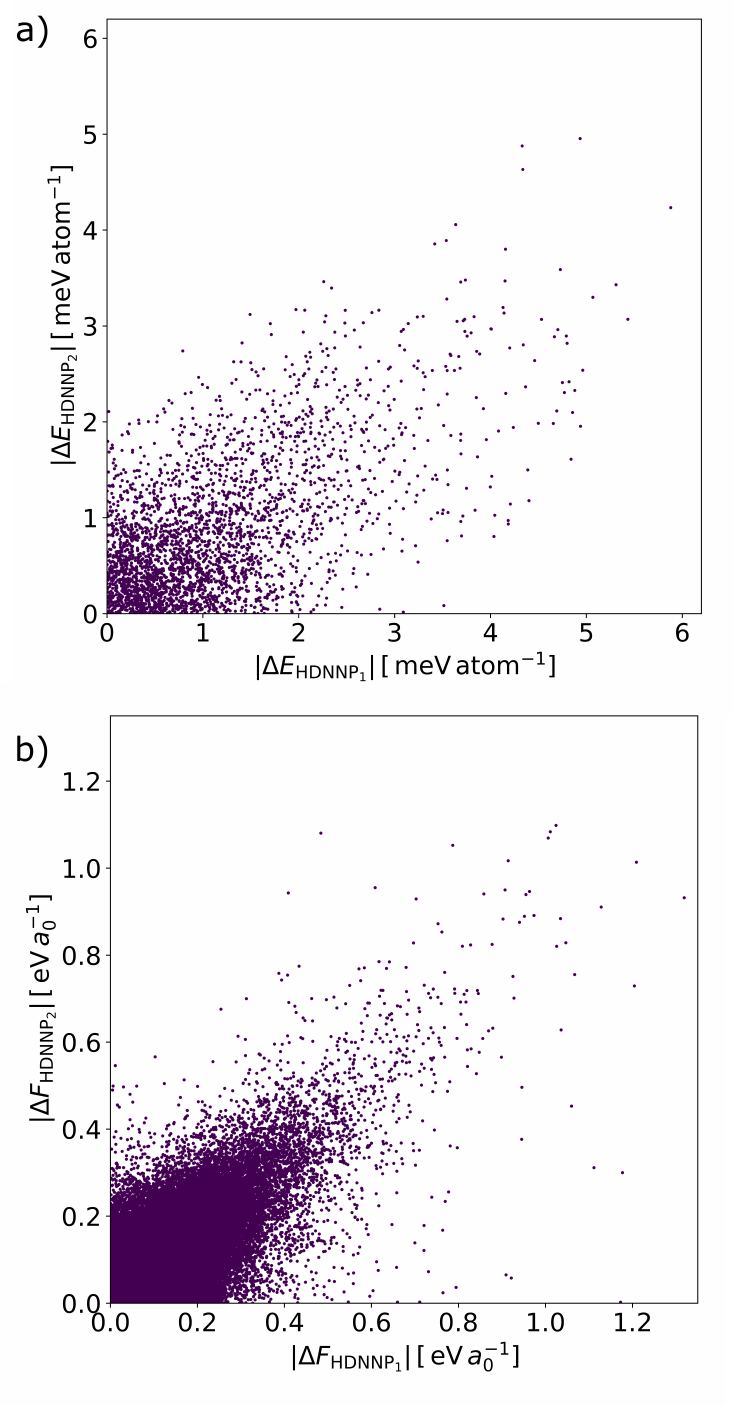}
    \caption{Error correlation for different HDNNPs fitted on the LiOH data set. Panels a) and b) show the correlation between the training set errors of total energies and atomic forces with respect to the reference data obtained with two different HDNNPs trained to the same data set. Data points with high errors in both potentials indicate possible contradictory data, which cannot be learned due to insufficient the accuracy of the reference data. Such problems might arise from underconverged settings in the reference calculations (cf. Fig.\ref{fig:grids})) or  incomplete electronic self-consistency.
    }
    \label{fig:NNP1_NNP2}
\end{figure}

Due to the absence of physically restricted functional forms in MLPs, the validation is of central importance not only for the final potentials but also during the training process and during the extension of the data set by active learning, as all these steps cannot be strictly separated. 
The validation is a multistep process, which is illustrated schematically in Fig.\ref{fig:multistepvalidation}. Starting from the available reference data set, a set of preliminary HDNNPs is constructed. For this purpose, the data set is split randomly into a training and a test set, and early-stopping is used to generate potentials with low test set RMSEs indicating good generalization of the potentials to structures not included in the training set. Like in the case of learning curves discussed above, the early-stopping method can provide information about the density and completeness of the data in the configuration space that is covered by the reference data, but no assessment of the quality for other types of structures is possible. 
\\
For all data in the training and the test sets, further analyses need to be done, since the RMSEs of the energies and forces provide only averaged information. Therefore, outliers and individual energies or forces, which are difficult to learn, might remain undetected when inspecting RMSE values only. Fig.~\ref{fig:enn_eref} shows the energy correlation plot relating the energies predicted by the HDNNP to the reference energies. For reasonable potentials all points should be located close to the line of perfect correlation. Such plots should be routinely investigated for the energies and forces in the training as well as in the test set for all generated HDNNPs. Outliers in these plots can often be related to problems in the underlying reference data, such as failed electronic convergence. 
\\
A closer investigation of outliers is possible when plotting the correlation of errors obtained with two different potentials (see Fig.\ref{fig:NNP1_NNP2}). Points, which have a low error in one of the two or even in both potentials are acceptable, since the representation quality of individual points may differ from HDNNP to HDNNP. If, however, a point has large errors in both potentials, this is an indication of contradictory data in the training set. Such contradictions may arise, e.g, from 
inconsistent k-point sets (see Sec.~\ref{sec:reference}) resulting in the assignment of different energies or forces to very similar structural features. Such points cannot be learned and can be identified in error correlation plots for closer examination.
It is important to note in this context that even a small number of problematic data points can result in very poor HDNNPs, because they can have a strong impact on the weight optimization thus affecting also the quality of the overall PES.


\begin{figure}[!ht]
    \centering
    \includegraphics[scale=0.80]{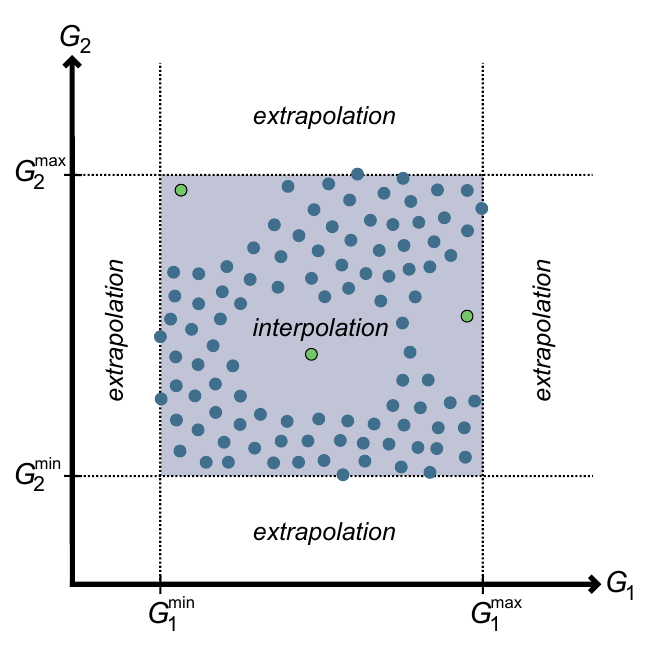}
    \caption{Illustration of poorly represented regions in a two-dimensional configuration space spanned by coordinates $G_1$ and $G_2$. The blue points represent the available training data covering the grey-shaded region. Any structure outside the minimum and maximum values of $G_1$ and $G_2$ is formally extrapolating. The three green structures, however, are inside the intervals $[G_1^{\mathrm{min}},G_1^{\mathrm{max}}]$ and $[G_2^{\mathrm{min}},G_2^{\mathrm{max}}]$ and therefore do not fulfill the criterion of extrapolation. Still, these structures are located in poorly represented regions and will be unreliable. Such structures can be identified by active learning.}
    \label{fig:interpolation}
\end{figure}

In the next step, a large number of configurations is generated, ideally by the simulation technique that shall be used in the production calculations. These configurations should be searched for extrapolations, i.e., the existence of ACSF values outside the range of function values defined by the training set (see Fig.~\ref{fig:interpolation}). This needs to be done separately for each ACSF, which is straightforward and computationally cost-effective, since information about the minimum and maximum values is typically available for each function from the preparatory scaling of the ACSFs before training. Since MLPs are not reliable in case of extrapolation, such situations must be avoided in the production simulations. This can be achieved by systematically searching for extrapolating structures to extend the reference data with the aim to cover an increasing part of configuration space. It should be noted, however, that the absence of extrapolation is not a sufficient criterion for a reliable potential, as structural outliers can drastically extend the range of symmetry function values without a suitable coverage of configuration space. In such a situation, extrapolation in the descriptor space cannot be automatically detected. Another form of extrapolation refers to the potential energy and forces in the system. Predictions of energies or forces outside the range of values in the training set should be carefully checked in production simulations, as they might be less reliable.
\\
In principle, extrapolating structures can also be found in the active learning process, which is the most general approach, since deviations in the energy and force predictions for different HDNNPs are very likely in this case. However, reasonable predictions even in the case of extrapolation cannot be excluded, and in such situations the elimination of extrapolating structures by active learning can be difficult. Once extrapolation occurs, simulations should be stopped and moderately extrapolating structures should be included, while structures exhibiting very different descriptor values should be discarded since very unphysical structures occuring in later stages of extrapolating trajectories should not enter the reference data set. As a simple check, for each newly added structure the interatomic distances should be computed and structures including too short bonds and other unwanted structural features should be excluded.
\\
Much more difficult than the detection of extrapolation is the identification of regions in configuration space, which are not sufficiently sampled, i.e., ``holes'' in the multidimensional data set. Examples for points in such holes, which are formally not fulfilling the criterion of extrapolation are shown in  Fig.~\ref{fig:interpolation}. These points can be identified by active learning as described above, which is thus a central component of potential validation. Only if the active learning process has been completed without finding further structures with substantial uncertainty, a reasonable transferability of the HDNNP can be expected.
\\
The last and most important quality check of a data set is the performance of the fitted HDNNP in applications. Whenever possible, direct comparisons with data obtained directly with the reference electronic structure method should be made, including e.g. equilibrium geometries and lattice parameters, vibrational frequencies and radial distribution functions. Ideally also independently trained HDNNPs, which have passed the full hierarchy of validation steps, should be employed to check on the robustness of the simulation results with respect to the specific parameterization.

\section{Conclusions}\label{sec:conclusion}

Much progress has been made in the past two decades in the construction of MLPs for atomistic simulations of large molecular and condensed systems. These developments do not only concern the general methodical framework, such as the adaption and use of modern machine learning algorithms and the derivation of suitable descriptors, but also the applicability to more and more complex systems. At the present time, the increasing availability of methods and their implementations in easily accessible software packages has substantially lowered the barrier to the  parameterization and use of MLPs. Still, this availability also bears some risks as the quality of MLPs is much more difficult to assess than the performance of conventional empirical potentials and force fields. 
\\
While MLPs in principle can reach a very high accuracy, which is often indistinguishable from the underlying electronic structure method, validation takes a central role as an excellent performance for some atomic configurations may go along with dramatic failures for other geometries. The use of ``canned'' potentials should therefore only be recommended if detailed information about the applicability of a potential for a specific system is available. Unfortunately, no community standards have been established for such information yet. 
\\
An important take-home message of this Tutorial is the insight that the parameterization, the generation of the reference data set and the validation of a potential are equally important steps, which cannot be separated from each other. An important example is active learning, in which preliminary, i.e., not yet perfect, potentials are used to identify structures missing in the training set. Hence, active learning also represents an important opportunity for validation on-the-fly. An entire hierarchy of validation steps is available, with active learning being the most general approach, which in principle also can cover the detection of overfitting and extrapolation. A lot has been achieved in (semi)automatic active learning schemes and the uncertainty quantification of MLPs during application, but the state of black-box methods has not yet been reached. Moreover, even some apparently simple tests, such as energy and force correlation plots, the analysis of emerging structures and the detailed inspection of outliers, which can often be traced back to some problem in the data set, can contribute substantially to the improvement of potentials. Although this Tutorial takes the perspective of high-dimensional neural network potentials, almost all discussed aspects are generally valid for a wide range of MLPs currently in use.

\begin{acknowledgments}
This work was supported by the Deutsche Forschungsgemeinschaft (DFG, German Research Foundation) under Germany’s Excellence Strategy—EXC 2033–390677874—RESOLV. Funding by the DFG (priority program SPP 2363, project number 495842446) and discussions with Marco Eckhoff are gratefully acknowledged. 
\end{acknowledgments}

\bibliography{literature}

\end{document}